\documentstyle[aaspp4]{article}

\newcommand{\etal}{{\sl et al.\/}}
\newcommand{\mpc}{{$h^{-1}$~Mpc\/}}
\newcommand{\sqdeg}{{$\mbox{deg}^2$\/}}
\def\ltsima{$\; \buildrel < \over \sim \;$}
\def\lsim{\lower.5ex\hbox{\ltsima}}
\def\gtsima{$\; \buildrel > \over \sim \;$}
\def\gsim{\lower.5ex\hbox{\gtsima}}

\begin{document}

\title{Detecting Clusters of Galaxies in the Sloan Digital Sky Survey I : Monte Carlo Comparison of Cluster Detection Algorithms}

\author{Rita Seung Jung Kim\altaffilmark{1}$^,$\altaffilmark{2} 
Jeremy V. Kepner\altaffilmark{1}$^,$\altaffilmark{3}, 
Marc Postman\altaffilmark{4}, Michael A. Strauss\altaffilmark{1}, 
Neta A. Bahcall\altaffilmark{1}, 
James E. Gunn\altaffilmark{1},  Robert H. Lupton\altaffilmark{1}, 
James Annis\altaffilmark{5}, 
Robert C. Nichol\altaffilmark{6}, 
Francisco J. Castander\altaffilmark{7}$^,$\altaffilmark{8}
J. Brinkmann\altaffilmark{9},
Robert J. Brunner\altaffilmark{10},
Andrew Connolly\altaffilmark{11},
Istvan Csabai\altaffilmark{2},
Robert B. Hindsley\altaffilmark{12},
\v{Z}eljko Ivezi\'{c}\altaffilmark{1}, 
Michael S. Vogeley\altaffilmark{13},
and Donald G. York\altaffilmark{14}}

\altaffiltext{1}{Princeton University Observatory, Princeton, NJ 08544, USA}
\altaffiltext{2}{Department of Physics and Astronomy, The Johns Hopkins University, 3701 San Martin Dr., Baltimore, MD 21218, USA}
\altaffiltext{3}{MIT Lincoln Laboratory, Lexington, MA, 02420, USA}
\altaffiltext{4}{Space Telescope Science Institute, 3700 San Martin Dr., Baltimore, MD 21218, USA}
\altaffiltext{5}{Fermilab, Batavia, IL 60510, USA}
\altaffiltext{6}{Department of Physics,  5000 Forbes Avenue, Carnegie Mellon University, Pittsburgh, PA 15213, USA}
\altaffiltext{7}{Yale University, P.O. Box 208101, New Haven, CT 06520-8101, USA}
\altaffiltext{8}{Universidad de Chile, Casilla 36-D, Santiago, Chile}
\altaffiltext{9}{Apache Point Observatory, P.O. Box 59, Sunspot NM 88349-0059, USA}
\altaffiltext{10}{Department of Astronomy, California Institute of Technology, Pasadena, CA 91125, USA}
\altaffiltext{11}{Department of Physics and Astronomy, University of Pittsburgh, Pittsburgh, PA 15260, USA}
\altaffiltext{12}{U.S. Naval Observatory, 3450 Massachusetts Ave., NW, Washington, DC 20392-5420, USA}
\altaffiltext{13}{Department of Physics, Drexel University, Philadelphia, PA 19104, USA}
\altaffiltext{14}{University of Chicago, Astronomy \& Astrophysics Center, 5640
S. Ellis Ave., Chicago, IL 60637, USA}

\begin{abstract}
We present a comparison of three cluster finding algorithms from 
imaging data using Monte Carlo simulations of clusters embedded in a 25
\sqdeg~region of Sloan Digital Sky Survey (SDSS) imaging data: the
Matched Filter (MF; Postman \etal~1996), the Adaptive Matched Filter
(AMF; Kepner \etal~1999) and a color-magnitude filtered Voronoi
Tessellation Technique (VTT).  Among the two matched filters, we find
that the MF is more efficient in detecting faint clusters, whereas the
AMF evaluates the redshifts and richnesses more accurately, therefore
suggesting a hybrid method (HMF) that combines the two.  The HMF
outperforms the VTT when using a background that is uniform, but it
is more sensitive to the presence of a non-uniform galaxy
background than is the VTT; this is due to the assumption of a uniform
background in the HMF model.  We thus find that for the detection
thresholds we determine to be appropriate for the SDSS data, the performance
of both algorithms are similar; we present the selection
function for each method evaluated with these thresholds as a function
of redshift and richness.  For simulated clusters generated with a
Schechter luminosity function ($M_r^* = -21.5$ and $\alpha = -1.1$)
both algorithms are complete for Abell richness $\gsim$ 1 clusters up
to $z \sim 0.4$ for a sample magnitude limited to $r=21$. 
While the cluster parameter
evaluation shows a mild correlation with the local background density, the
detection efficiency is not significantly affected by the
background fluctuations, unlike previous shallower surveys.

\end{abstract}


\section{Introduction}
\label{sec:intro}
Over the past few decades, clusters of galaxies have been used as
valuable tools for cosmological studies: they are tracers of large
scale structure (Bahcall 1988, Huchra \etal~1990, Nichol \etal~1992,
Nichol \etal~1994, Peacock \& Dodds~1994, Collins \etal~2000), their
number density is a constraint on cosmological models (Gunn \&
Oke~1975, Hoessel, Gunn \& Thuan~1980, Evrard~1989, Bahcall
\etal~1997, Carlberg \etal~1997, Oukbir \& Blanchard~1997, Reichart
\etal~1999), and they act as laboratories for probing the formation
and evolution of galaxies and their morphologies (Dressler~1980,
Butcher \& Oemler~1984, Gunn \& Dressler~1988, Dressler \etal~1997).
While it is sometimes sufficient simply to have a large sample of
clusters, most cosmological studies require a homogeneous sample with
accurate understanding of the selection biases and the completeness of
the catalog.

The Abell cluster catalog is by far the most widely used
catalog to date (Abell 1958; Abell \etal~1989, hereafter ACO).  Like
some of the other cluster catalogs that are available (e.g., Zwicky
\etal~1968; Gunn \etal~1986), the Abell catalog was constructed
entirely by visual inspection of photographic plates.  Although the
human eye is a sophisticated and efficient
detector for galaxy clusters, it suffers from subjectivity and
incompleteness, and visual inspection is extremely time consuming. For 
cosmological studies, the major disadvantage of such visually
constructed catalogs is that it is difficult to quantify selection
biases and the selection function.

The main motivation for automated cluster finding schemes is thus to
overcome these major drawbacks of visual catalogs, namely speed,
objectivity and reproducibility. The first automated cluster finding
in optical surveys was attempted by Shectman (1985), and was followed
by several variants of this peak finding method; Lumsden \etal~(1992)
with the EDSGC (Nichol, Collins, \& Lumsden~2001) and Dalton
\etal~(1992) with the APM survey (Maddox \etal~1990).  Lahav \& Gull
(1989) introduced the concept of fitting the observed apparent
diameter distribution to the absolute distribution to obtain an
estimate of the distance to each cluster, which led to the development
of the Matched Filter Algorithm (Postman \etal~1996, hereafter
P96). The matched filter technique has been widely used, and several
variants have been further developed (Kawasaki \etal~1998; Schuecker \&
B\"{o}hringer 1998; Lobo \etal~2000), including the Adaptive Matched
Filter (Kepner \etal~1999; hereafter K99, Kepner \& Kim 2000).  In
addition, thanks to the availability of multi-band CCD imaging data,
several cluster finding methods have been put forward that utilize the
color characteristics of galaxy clusters (Annis \etal~2001, Gladders
\& Yee~2000, Goto \etal~2001, Nichol \etal~2000b).  While these methods can be
efficient, each inevitably has its own biases, depending on the
nature and extent of constraints that have been imposed for the
selection.  Little effort has been made to date to compare the
performances of these different methods.

In this paper, we present a comparison of three different cluster
finding algorithms, using a Monte Carlo experiment with simulated
clusters. The algorithms we analyze are the Matched Filter algorithm
(P96), the Adaptive Matched Filter algorithm (K99), and the Voronoi
Tessellation Technique, which is introduced in detail in
\S\ref{sec:vtt}.  The last technique, which uses color information, is
in part based on previous cluster finding efforts that use the
classical Voronoi Tessellation as a peak finder (Ramella \etal~1998,
Ebeling~1993, Kim \etal~2000), but is introduced here for the first
time in its current form. These three methods constitute the basis of
a cluster catalog derived from 150 \sqdeg~of Sloan Digital Sky Survey
(York \etal~2000; hereafter SDSS) commissioning data, which we present
in Paper II (Kim \etal~2001).  Work in progress (Bahcall \etal~2001)
will present cluster catalogs using a wide range of finding
techniques.  This paper is thus geared towards understanding the
behaviour of the cluster finding algorithms in the SDSS data.

In \S2 we describe the various cluster detection algorithms, and in
\S3 we present the Monte Carlo experiment in which these cluster
finding algorithms were run. The results and comparison between
methods are presented in \S4.  We summarize in \S5.  Throughout this
paper we assume a cosmology in which $\Omega_m = 0.3$,
$\Omega_{\Lambda}=0.7$ and $H_0 = 70 \mbox{km}\,\mbox{s}^{-1}\, \mbox{Mpc}^{-1}$ unless noted otherwise.


\section{Cluster Selection Algorithms}
The Matched Filter algorithm, which was first presented as a fully
automated cluster finding scheme in P96, has been widely adapted for a
variety of cluster detection efforts (e.g., Olsen \etal~1999,
Scodeggio \etal~1999, Bramel \etal~2000; hereafter BNP00, Postman
\etal~2001, Willick \etal~2001).  The Adaptive Matched Filter
algorithm (K99, Kepner \& Kim 2000), which is described below, should
be a substantial improvement both in content and efficiency over the
original Matched Filter algorithm; the major changes being the
adoption of a full likelihood function and the incorporation of 3-D
(redshift) information.  Nevertheless, unlike the P96 Matched Filter,
the Adaptive Matched Filter algorithm has not yet been applied to real
data, and therefore lacks the optimizations and adaptations that
exposure to real data would give.  Thus we have chosen to apply both
the original Matched Filter (hereafter MF) and the Adaptive Matched
Filter (hereafter AMF) for comparison and cross-checking
purposes. Although the AMF should in theory converge to the MF results
in the 2-D case, there are various differences in the details of the
two codes (e.g., peak selection criteria, final parameter evaluations,
likelihood function detailed further below) that can cause the final
results to differ somewhat.  The fact that they are two completely
independent codes written in two different languages (C and IDL) also
makes the cross check particularly useful.

The Matched Filter technique is an efficient likelihood method for
finding clusters in two dimensional imaging data.  A model cluster
radial profile and galaxy luminosity function are used to construct a
matched filter in position and magnitude space from which a cluster
likelihood map is generated. Using the magnitudes rather than simply
searching for density enhancements suppresses false detections that
occur by chance projection. Of course, the results are dependent on
the assumed filter shape and extent; they are thus affected by the assumptions
made for the universal radial profile and cluster luminosity function
used in the algorithm.  In other words, the cluster parameters that
are derived from the algorithm are somewhat dependent on the model
that was assumed for the cluster, and the probability for detecting a
certain cluster may differ as the cluster shape varies (e.g.,
spherical vs. elongated), or as the cluster parameters deviate from
the assumed cluster model.  In fact most clusters {\it are} elongated,
and are known to have a variety of shapes (e.g., Bautz-Morgan (1970)
type, Rood-Sastry (1971) type).  The third method, the Voronoi
Tessellation Technique (hereafter VTT), was in part motivated by this
model-dependent aspect of the Matched Filter algorithms (MF and AMF),
to examine if any severe biases occur in the selection due to model
assumptions.  In addition to the fact that the VTT assumes no
intrinsic cluster properties (except for very mild constraints in
color-magnitude space, see below), it is simple and fast (for $n=10^6$
galaxies, it takes seconds on a 400~MHz CPU to evaluate Voronoi
Tessellation).

The details of the MF and the AMF are given in the respective
references, so here we only describe them briefly
(\S\ref{sec:mf}), mainly comparing them carefully and
emphasizing their differences. This is followed by a detailed recipe for
the VTT in \S\ref{sec:vtt}.

\subsection{The Matched Filter Techniques}
\label{sec:mf}
The foundation for both matched filter techniques is the model for
the total number of galaxies per unit area per unit observed flux $l$:
\begin{equation}
n_{\mbox{model}} (\theta,l,z_c)\, dA\, dl = [ n_f(l) + 
	\Lambda_{cl} n_c (\theta,l,z_c)]\, dA \,dl ,
\label{eq:nmodel}
\end{equation}
which consists of contributions from background field galaxies
($n_f$) and the cluster galaxies ($\Lambda_{cl}n_c$) at redshift
$z_c$, where $\Lambda_{cl}$ is the cluster richness measure (see below).  
Here, $\theta$ is the angular distance from the
cluster center, and $dA = 2\pi\theta \, d\theta$. The background
number density ($n_f$) is simply taken from the global number counts
of the survey. The essential ingredient of the matched filter is the
model for the cluster number density, $\Lambda_{cl} n_c$, which is a
product of a projected cluster density profile and a luminosity
profile for a cluster at redshift $z_c$:
\begin{equation}
\Lambda_{cl} n_c(\theta,l;z_c) = \Lambda_{cl} \Sigma_c(r) \left( {dr\over d\theta} \right)^2
\phi_c (L) \left( {dL \over dl } \right) ,
\end{equation}
where $r$ is the projected comoving radius and $L$ is the absolute
luminosity. The conversion from physical units to apparent units
includes proper treatment for cosmology and the K-correction. For the
projected density profile we use the modified Plummer law model
(see K99 Appendix A),
\begin{equation}
\label{eq:plummer}
\Sigma_c(r) \propto
\left\{ 
\begin{array}{l}
[1+(r/r_c)^2]^{-n/2} - [1+(r_{\mbox{max}}/r_c)^2]^{-n/2}  \ \ \ \  \mbox{for} \  (r<r_{\mbox{max}}) \\
 0   \ \ \ \ \mbox{for} \  (r \geq r_{\mbox{max}})
\end{array}
\right.
\end{equation}
where $n \approx 2$, the slope of the profile, $r_c$ is the core
radius and $r_{\mbox{max}}$ is a cutoff radius which approximates the
extent of a cluster. This cutoff radius naturally constrains the
radius of the search; forcing $\Sigma_c(r_{\mbox{max}}) = 0$ is
equivalent to putting a ``taper'' on the filter which reduces the
``sidelobes'' and narrows the ``beam''.  This is standard practice in
detection theory, which improves the spatial accuracy and makes the
process more robust (at the cost of slightly reducing the sensitivity
of the filter), as well as reducing the contamination that arises from
other nearby clusters.

Any method of smoothing the data consists of choosing a filter shape
and a filter bandwidth. Numerical experiments show that the efficiency
of the estimator is much more sensitive to the filter bandwidth than
is the filter shape itself (Silverman~1986). Hence determining the
appropriate values of $r_{\mbox{max}}$ and $r_c$ is more important
than our particular choice of the cluster profile functional form.
P96 (see \S4 in their paper) discuss the effect that the cutoff radius
has on their detection efficiency and conclude $r_{\mbox{max}} =
1$\mpc~as an optimal choice, which we adopt as an appropriate value.
Increasing the value of $r_{\mbox{max}}$ further will significantly
degrade the signal-to-noise ratio; the cluster signal will go down
since less weight is given to the core, and noise from the background
and nearby clusters will increase as more weight is given to galaxies
at larger radii.

For the luminosity profile we adopt a standard Schechter luminosity
function (Schechter 1976),
\begin{equation}
\phi_c(L)dL \propto (L/L^*)^{\alpha} e^{-L/L^*} d(L/L^*) .
\end{equation}
The overall normalizations of $\Sigma_c$ and $\phi_c$ are
chosen such that the cluster richness measure $\Lambda_{cl}$ is the
total cluster luminosity within $r_{\mbox{max}}$ 
in units of $L^*$, i.e., $L_{cl} (\leq r_{\mbox{max}}) = \Lambda_{cl} L^*$
(see P96 and K99 for details).  The Abell richness $N_A$ is defined by
the number of cluster galaxies (within $r < 1.5$\mpc) with magnitudes
between $m_3$ and $m_3 + 2$, where $m_3$ is the magnitude of the third
brightest galaxy in the cluster.  The Abell Richness Class (RC;
Abell~1958) is determined by this quantity $N_A$; $30 \leq N_A \leq
49$ for RC = 0, $50 \leq N_A \leq 79$ for RC = 1, $80 \leq N_A \leq 129$
for RC = 2, and $130 \leq N_A \leq 199$ for RC = 3.   The relation between
$N_A$ and $\Lambda_{cl}$ (with $r_{max} = 1$\mpc) is found to be
approximately $N_A \sim (2/3) \Lambda_{cl}$ (Bahcall \& Cen 1993, P96)
but with large scatter (P96).
This relation is addressed further in Paper II with the clusters
detected from the Sloan Digital Sky Survey imaging data.

The difference between the MF and the AMF starts from the definition
of the likelihood function. The MF adopts a Gaussian likelihood
function, which is based on the assumption that there are enough
galaxies in each virtual bin ($j$) in (position, magnitude) space
that the Poisson probability distribution can be approximated by a
Gaussian. Furthermore, it assumes that the background galaxy
distribution ($n_f$) is uniform and large enough to dominate the
noise, therefore the likelihood can be written as,
\begin{equation}
{\cal L}(\theta) = \sum_j -2 \ln P_{j} = -2 \sum_j \ln 
\left[ {1 \over \sqrt{2 \pi n_f^{(j)} }}
\exp \left( 
- {(n^{(j)}_{\mbox{data}} - n^{(j)}_{\mbox{model}}(\theta))^2 \over 2n_f^{(j)}}
\right)
\right] .
\end{equation}
By approximating the summation with an integral, using
Eq.~(\ref{eq:nmodel}) for $n_{\mbox{model}}$, setting $\delta \equiv
n_c/n_f$, and dropping all terms irrelevant to $\theta$, one obtains a
likelihood function that is linear in the data and easy to calculate:
\begin{equation}
{\cal L} = \Lambda_{cl}\sum_i \delta_i , \ \ \ \ \mbox{where} \ \ \ 
\Lambda_{cl} = {    \sum_i \delta_i \over 
	\int \delta(\theta,l) n_c(\theta,l) dA dl} ,
\label{eq:lcoarse}
\end{equation}
where $i$ stands for each galaxy, and the sum is over every galaxy
within $r_{\mbox{max}}$. The richness measure $\Lambda_{cl}$ is
obtained by first solving the equation $\partial {\cal L} /
\partial\Lambda_{cl} = 0$ (see K99 Appendix C or P96 for details).

In the AMF, this likelihood function is referred to as the ``coarse
filter''.  Since it is simple and easy to calculate, AMF uses
Eq.~(\ref{eq:lcoarse}) to construct a ``coarse'' likelihood map in
order to select clusters, by identifying peaks from this map. However,
the Gaussian approximation breaks down when there are not enough
galaxies, i.e., especially for poor clusters or those at high
redshift, and in general the parameters that are evaluated by this
``coarse filter'' are found to be somewhat biased (P96 discuss
empirical correction factors for this bias, which we will not discuss
here in detail).  Hence, the AMF defines a second likelihood function
that assumes a Poisson likelihood instead of a Gaussian (Dalton
\etal~1994).  This is called the ``fine filter'' and reduces to:
\begin{equation}
{\cal L}_{\mbox{fine}} = -\Lambda_{cl} N_c + \sum_i \ln (1+\Lambda_{cl}
\delta_i) ,
\end{equation}
where $\Lambda_{cl}$ is obtained by solving 
\begin{equation}
\label{eq:nc}
N_c = \sum_i {\delta_i \over 1+\Lambda_{cl} \delta_i} .
\end{equation}
$N_c$ is the number of galaxies expected in a $\Lambda_{cl} = 1$
cluster (see K99 Appendix C for the derivation). The resulting fine
filter likelihood function is nonlinear and requires more computations
to evaluate. However, as it is based on correct statistics, the
evaluated cluster parameters (redshift and richness) are expected to
be more accurate.  The AMF thus adopts a two-layered approach, first
to identify clusters by peaks in the coarse filter likelihood map, and
then to evaluate proper parameters by the fine filter on those
selected cluster positions.  The computing time difference between the coarse
and the fine filter is due to solving Eq.~(\ref{eq:nc}), which must be
done at every grid position; using only the fine filter would take
approximately 10 times longer.  In addition, the AMF approach allows
an internal cross check for the evaluation of $z$ and $\Lambda_{cl}$.

The MF and AMF have further differences that turn out, as we will
see, to be quite important.  The MF uses a uniform grid in 2D space
on which the likelihood function is evaluated for a series of redshift
values that span the desired redshift range for the cluster search.
The grid size is a function of redshift assumed; we use a grid size of
1/2 the core radius at each redshift to ensure proper sampling.  Hence the
result is a series of likelihood maps evaluated for each assumed
redshift, i.e., a map indicating the likelihood for
the existence of a cluster at that redshift.  The AMF instead uses a
so-called ``naturally adaptive grid'', evaluating the likelihood
function at the galaxy positions themselves, meaning that as we go to
higher redshift, the effective resolution becomes finer as necessary.
Such a non-uniform grid is slightly more complicated to handle, 
but ensures that a unique
galaxy lies at the center of each cluster, and no computation is wasted on
unnecessary positions.  As a result, instead of
producing series of likelihood maps to be stored at each assumed
redshift value, the AMF calculates the likelihood values on a redshift
grid for each galaxy separately, and records the redshift and
likelihood at which the likelihood is maximized.  In other words, the
outcome is two quantities at each galaxy: the peak
likelihood that a cluster lies at this point, and the corresponding
estimated redshift.
This saves a significant amount of disk space relative to the MF, and
is cleverly structured to reduce an intermediate step in the analysis
(step 5 in Table~\ref{tab:mf}), but as we shall see below, this
approach has drawbacks.
 
Once the coarse likelihood map is generated, the two algorithms differ
in the final steps of the cluster selection process. The MF, having
stored likelihood maps for each redshift bin, finds local maxima in
each map, and registers them as candidate clusters when they lie above
a prescribed threshold for each map (e.g., 95 percentile within the
map; approximately $2 \sigma$ level).  This is repeated for every
redshift bin, then all the cluster candidates from all redshift bins
are combined together to filter out overlaps and to find the most
likely redshift for each cluster -- the redshift where the peak signal
is maximized.  Each cluster also has a significance of detection
$\sigma_{\mbox{det}}$ which is translated from the pixel distribution
of the likelihood signals at the final redshift assuming Gaussian
statistics.  We perform
a final cut on this quantity ($\sigma_{\mbox{det}} \geq
\sigma_{\mbox{cut}}$), to select significant detections.  On the other
hand, the AMF has already stored the most likely redshift of each
galaxy point, and therefore simply locates the position of the highest
likelihood signal, registers it as a cluster, eliminates all galaxies
around this point within a sphere of a given cluster size
($r_{\mbox{max}}$), and then looks for the next highest peak, and so
on until the likelihood signal (${\cal L}_{\mbox{coarse}}$) drops
below a prescribed threshold (${\cal L}_{\mbox{cut}}$).  The
difference between the two algorithms is not merely the order in which
the procedures are executed (the MF locates clusters in angular space
first and then determines the redshift, while the AMF determines the
redshift first for every point in space and then filters out the
clusters in angular space), but that the AMF uses a threshold in 
likelihood that is constant, thus redshift independent, while the MF
uses a threshold that differs for each redshift.
The effect that is caused by these differences is
discussed in \S\ref{sec:discussion}.

Finally, the AMF is completed by
evaluating the fine filter on the cluster positions that have been
selected, then determining the redshift at which the peak of the
fine likelihood occurred, and calculating the richness $\Lambda_{cl}$
for that redshift.  Table~1 summarizes the procedures and the parameters 
outlined above for the two Matched Filter algorithms.

\subsection{The Voronoi Tessellation Technique}
\label{sec:vtt}

The Voronoi Tessellation made its {\it debut} in astrophysics as a
convenient way of modeling the large-scale structure of the universe
(Icke \& van de Weygaert 1987, Ling 1987).  With a distribution of
seeds (nuclei), Voronoi Tessellation creates polyhedral cells
that contain one seed each, enclosing all the volume that is closest
to its seed. This is the definition of a Voronoi cell.  This natural
partitioning of space by Voronoi Tessellation has been used to model
the large scale distribution of galaxies.  This is achieved by
envisioning the seeds to be the expansion centers of ``voids'', the
planes that intersect two adjacent cells as ``walls'', the ridges
where three walls intersect as ``filaments'', and the vertices where
four filaments come together as galaxy clusters (van de Weygaert \&
Icke 1989).

A slightly different application of the Voronoi Tessellation is to
identify X-ray sources by locating the overdensity in X-ray photon
counts (Ebeling 1993, Ebeling \& Wiedenmann 1993); this is directly related to our application for
cluster finding, as we now describe. The galaxy positions are input as
the seeds for the Voronoi Tessellation, and the Voronoi cell around
each galaxy is interpreted as the effective area that each galaxy
occupies in space.  Taking the inverse of these areas gives a local
density at each galaxy in two dimensions.  This information is then
used to threshold and select galaxy members that live in highly
over-dense regions, which we identify as clusters.  We do so by
calculating the density contrast at each galaxy position $\delta
\equiv (\rho - \bar{\rho})/\bar{\rho} = (\bar{A} - A)/A $, where $A$
is the area of the Voronoi cells, and $\bar{A}$ is the mean area of
all cells.  We then impose a cut in the density contrast $\delta >
\delta_{c}$ to select galaxies in high density environments.  One can,
in fact, use a more rigorous statistical approach for the detection
criteria, using statistics of Voronoi Tessellation for a random
distribution of seeds (Kiang 1966; see Ramella \etal~1999, 2001 for
details).  However, as our approach described below is empirical, we
adhere to a simple cut in constant density contrast whose value is
tested by a Monte Carlo method using simulated clusters, as described
in section~\ref{sec:montecarlo}.

The SDSS is currently working on determining
photometric redshifts for galaxies in the imaging data,
but until they are available, we are confined to working with the two
dimensional projected distribution; therefore we need to divide the galaxy
sample to group them into comparable redshifts in order to enhance the
cluster detectability.  Therefore, our recipe for the Voronoi
Tessellation Technique utilizes {\it a priori} knowledge of
characteristics of cluster member galaxies, namely, the
color-magnitude relation.

\begin{figure}	
\plotfiddle{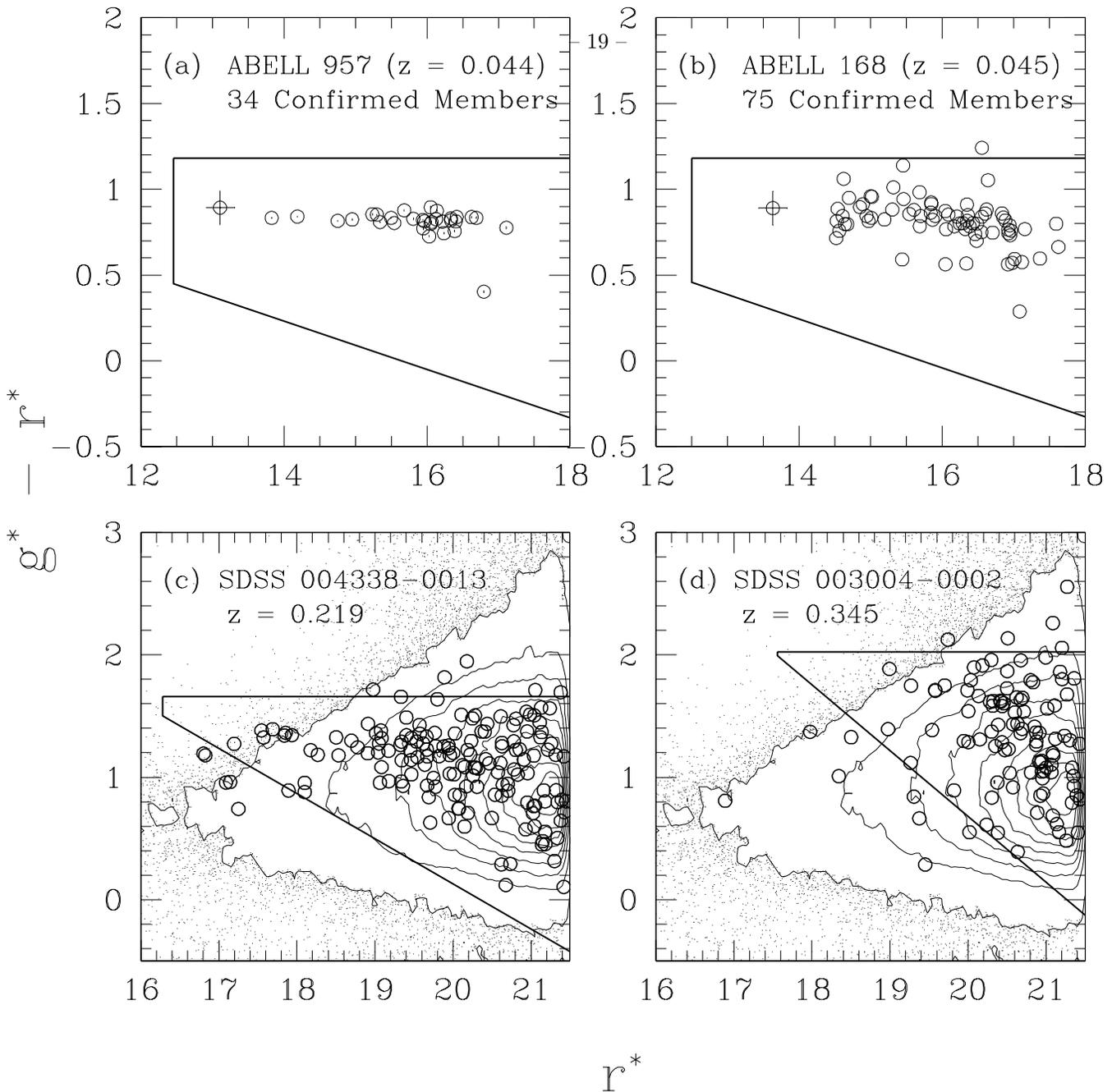}{5.8in}{0}{90}{90}{-290}{-140}
\caption{Color-magnitude (C-M) diagrams of four different clusters.
(a) and (b) are known Abell clusters, and plotted here are the data
from only those galaxies with confirmed membership according to the
ENACS (Katgert \etal~1998).  All data are taken from SDSS
photometry.
(c) and (d) show C-M diagrams of new clusters found in the
SDSS itself by the matched filter techniques.
Both clusters were confirmed visually through 3 color
($g,r,i$) composite images and their redshifts ($z=0.219$ and
$z=0.345$) were obtained by the SDSS spectroscopic survey (York
\etal~2000). The circles represent galaxies within 1 $h_{70}^{-1}$Mpc
of the detected center ($r=4.74'$ and $r=3.45'$) and the contours
represent the C-M distribution of all galaxies from a 25 \sqdeg~region
around the cluster.  The thick solid lines in all four panels show the C-M
filtering for the VTT (see text) at the corresponding redshifts,
enclosing the region which most cluster galaxies inhabit for a given
redshift. $g^*-r^*$ colors are from model magnitudes and $r^*$ is in
Petrosian magnitudes.
\label{fig:cmdiag}}
\end{figure}

Galaxies within a cluster usually exhibit a tight correlation between
their colors and magnitudes. It is well known that the core of a
typical rich cluster consists mainly of early type galaxies (i.e.,
Hubble Types E and S0; Hubble 1936, Oemler 1974, Postman \& Geller
1984, Dressler~1980, 1984) which all have very similar red colors.
This includes the Brightest Cluster Galaxy (hereafter BCG), whose
properties have been well studied (Schneider, Gunn \& Hoessel~1983,
Postman \& Lauer 1995).  Figures~\ref{fig:cmdiag}a~and~b show the
color-magnitude relation (C-M diagram) of Abell Clusters 168 and 295
respectively, observed with the SDSS camera; the diagram shows only those 
galaxies whose cluster membership has been confirmed spectroscopically
by the ENACS survey (Katgert \etal~1998).  The BCG is marked with a
cross, and the narrow horizontal line of galaxies at nearly constant color
is referred to as the E/S0 ridgeline (Visvanathan \& Sandage 1977,
Annis \etal~1999).  The color-magnitude relation for E/S0 galaxies
has been well known since Baum (1959), who first noted
that fainter early type galaxies tend to be bluer, showing a
negative slope of the E/S0 ridgeline in C-M diagrams (see also Visvanathan \&
Sandage 1977, Lugger 1984, Metcalfe \etal~1994).  It has been
suggested that this slope evolves with redshift (Gladders \etal~1998)
and even differs with richness (Stoughton \etal~1998). However, as we
see in the C-M diagrams, the slope is very shallow in $(g^*-r^*)$ vs. $r^*$ 
space, and is a negligible effect for the recipe that we now describe.

We will use this characteristic shape in the C-M diagram to select against
foreground and background galaxies.  Thus we carry out the following
approach: first select a redshift, then apply a Voronoi Tessellation
on all galaxies in a restricted region of the C-M diagram. This region
is shown as solid lines in Figures~\ref{fig:cmdiag}a~and~b, for
a redshift of $z=0.045$, whose limits enclose most of
the galaxies that are confirmed members of the cluster.  
Figures~\ref{fig:cmdiag}c~and~d show similar C-M diagrams of cluster
candidates at higher redshifts, found by both the MF and the AMF from
the SDSS. These clusters were visually confirmed with SDSS images, and the  
redshifts of their BCGs were obtained by the SDSS spectroscopic
survey (York \etal~2000).  Since membership information is
not available, we simply plot all the galaxies that are within
1\mpc~radius from the cluster center as circles.  For comparison, the
C-M distribution of all the galaxies in a survey region of 25 \sqdeg~is
shown in contours and small dots.  The difference in these
distributions illustrates the efficiency of using C-M information in
discriminating cluster members against the background population. Here
again, the areas enclosed by the solid lines are those that are
selected for detecting a cluster at $z=0.22$~and~$0.35$
respectively, which includes most of the galaxies within
$r=1$\mpc~around the cluster center.

The empirical C-M limits we use are as follows:
\begin{eqnarray}
\label{eq:cmlimits1}
  r^*_{bcg} - 1  <&  r^* &<  r^*_{bcg} + 5 \\
   a r^* + b  <& (g^*-r^*) &< (g^*-r^*)_{bcg} + 0.3 
\end{eqnarray}
where $a$ and $b$ are given by a simple linear relation with redshift, 
\begin{equation}
\label{eq:cmlimits2}
 a = -1.295 z - 0.084 , \ \ \ \  b = 30.13 z + 0.88
\end{equation}
where $r^*_{bcg}$ and $(g^*-r^*)_{bcg}$ are the Petrosian $r^*$ magnitude
and the model $g^*-r^*$ color of the BCG, respectively (see
\S\ref{sec:data} for description of magnitudes and colors).  These
relations are established empirically, by examining the C-M relations
of known clusters (e.g., Fig.~\ref{fig:cmdiag} a,b) and rich clusters
of galaxies found by the Matched Filter algorithms and the MaxBCG
technique (Annis \etal~2001) from the SDSS data itself.  The limits
are chosen to generously include most of the galaxies within 1\mpc~of
the cluster center.  The lower limit in magnitude, 5 magnitudes
fainter than that of the BCG, is chosen to cover the magnitude
range of spectroscopically confirmed cluster members from the ENACS
(Katgert \etal~1998) for low redshift clusters ($z < 0.1$). This is a
moderate coverage of cluster galaxies for a typical cluster luminosity
function (corresponding to $M_{lim} \sim L^* + 3$); also note the
recent findings of fossil groups whose difference between the BCG and
the second luminous galaxy can be as large as 4-5 magnitudes
(Zabludoff \& Mulchaey~1998). 

\begin{figure}
\plotfiddle{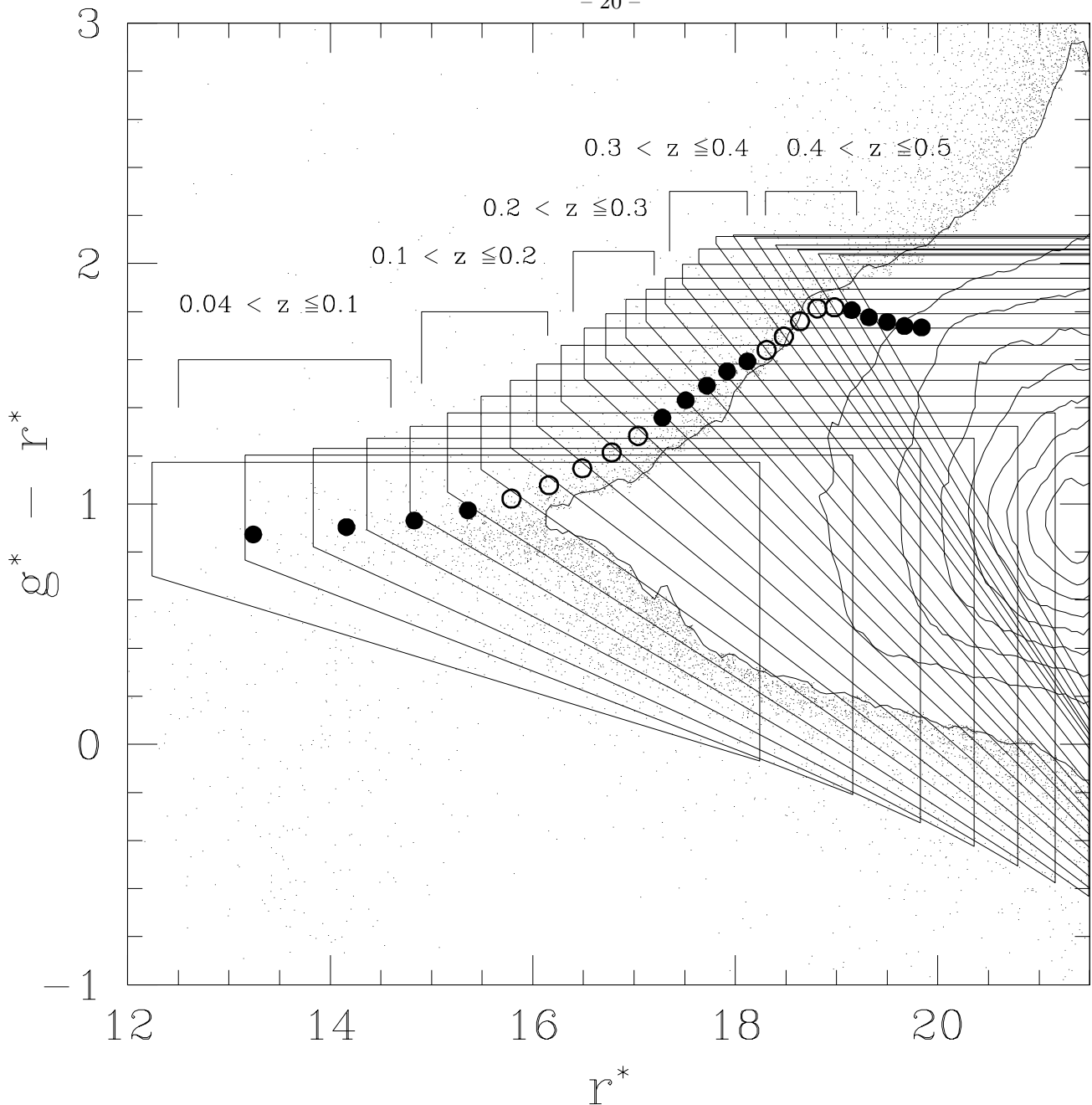}{5.8in}{0}{90}{90}{-290}{-140}	
\caption{The empirical color-magnitude limits used in the Voronoi
Tessellation Technique displayed for a range of redshifts ($0.04 \leq
z \leq 0.5, \Delta z = 0.02$). These limits are used as a filter to
enhance the signal of clusters at a given redshift (see text and
Eq.~(\ref{eq:cmlimits1}) \& (\ref{eq:cmlimits2})).  The large dots
trace the evolutionary track of a bright red galaxy with a constant
luminosity $M_{r^*} = -23$, intended to represent the Brightest Cluster
Galaxy.  Each C-M limit encloses one BCG, and the redshift range of
the C-M limits are labeled. The dots alternate as filled and
open for each redshift range labeled, for easy identification.
The contours and the small dots show the C-M distribution of all galaxies
in the SDSS survey (taken from a 150 \sqdeg~region of the SDSS
commissioning data) for comparison.
\label{fig:cmlimits}}
\end{figure}

These C-M limits are shown in summary for a range of redshifts, from
$z=0.04$ to $z=0.5$, in Figure~\ref{fig:cmlimits}.  Here, the track of
BCGs for the above redshifts are also shown as large dots (filled and
open), which provides the basis for determining these limits for each
redshift (Eqs.~(\ref{eq:cmlimits1})~and~(\ref{eq:cmlimits2})).
Finally, the C-M distribution of all galaxies in the survey region are
shown as contours and small dots for comparison.  The BCG track is
computed assuming a constant absolute magnitude $M(r^*)_{bcg} = -23$
with no evolution (see Eisenstein \etal~2001).  We use the PEGASE
evolutionary synthesis model (Fioc \& Rocca-Volmerange~1997) to
generate the spectral energy distribution of the BCG; this is then
redshifted and convolved with the SDSS filter responses to obtain
color K-corrections.

\begin{figure}
\plotfiddle{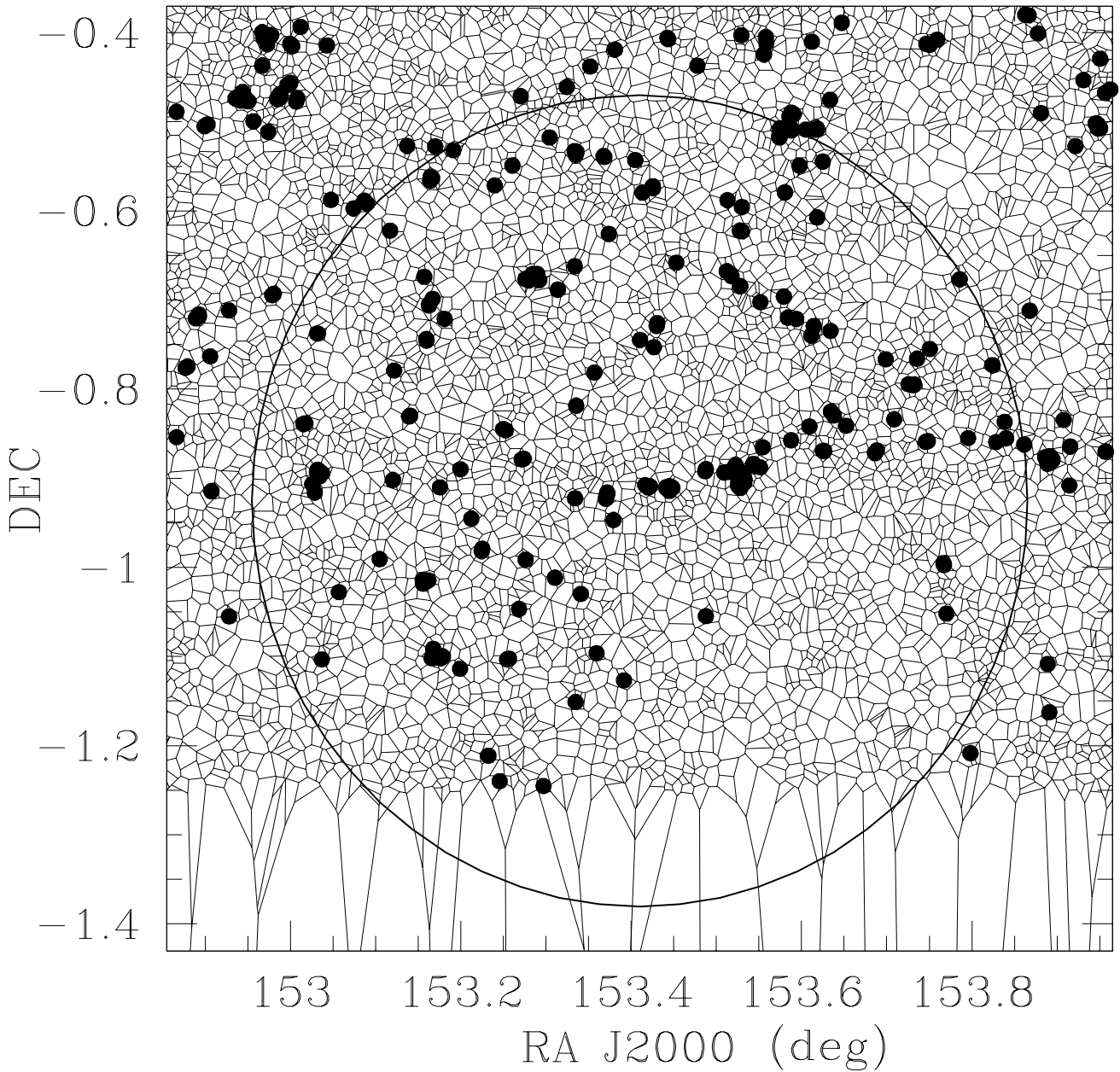}{5.8in}{0}{90}{90}{-290}{-140}	
\caption{ Example of Voronoi Tessellation executed on the galaxy
distribution around Abell cluster 957 (see Figure~1(a)). These are all
galaxies with $r^* < 21$. Each cell encloses one galaxy. The data
presented here has a lower boundary in declination of
$\delta_{\mbox{J2000}} = -1.25^{\circ}$, which is why the Voronoi
Tessellation seems to diverge below. The filled dots mark galaxies
with $\delta > 3$ (see text), and the large circle has a radius of
1\mpc~at a redshift of $z = 0.044$ ($r=26.4'$). We do not find any
significant overdensity of the filled dots around Abell 957 when using
the entire distribution of galaxies.
\label{fig:vttplot1}}
\end{figure}

\begin{figure}
\plotfiddle{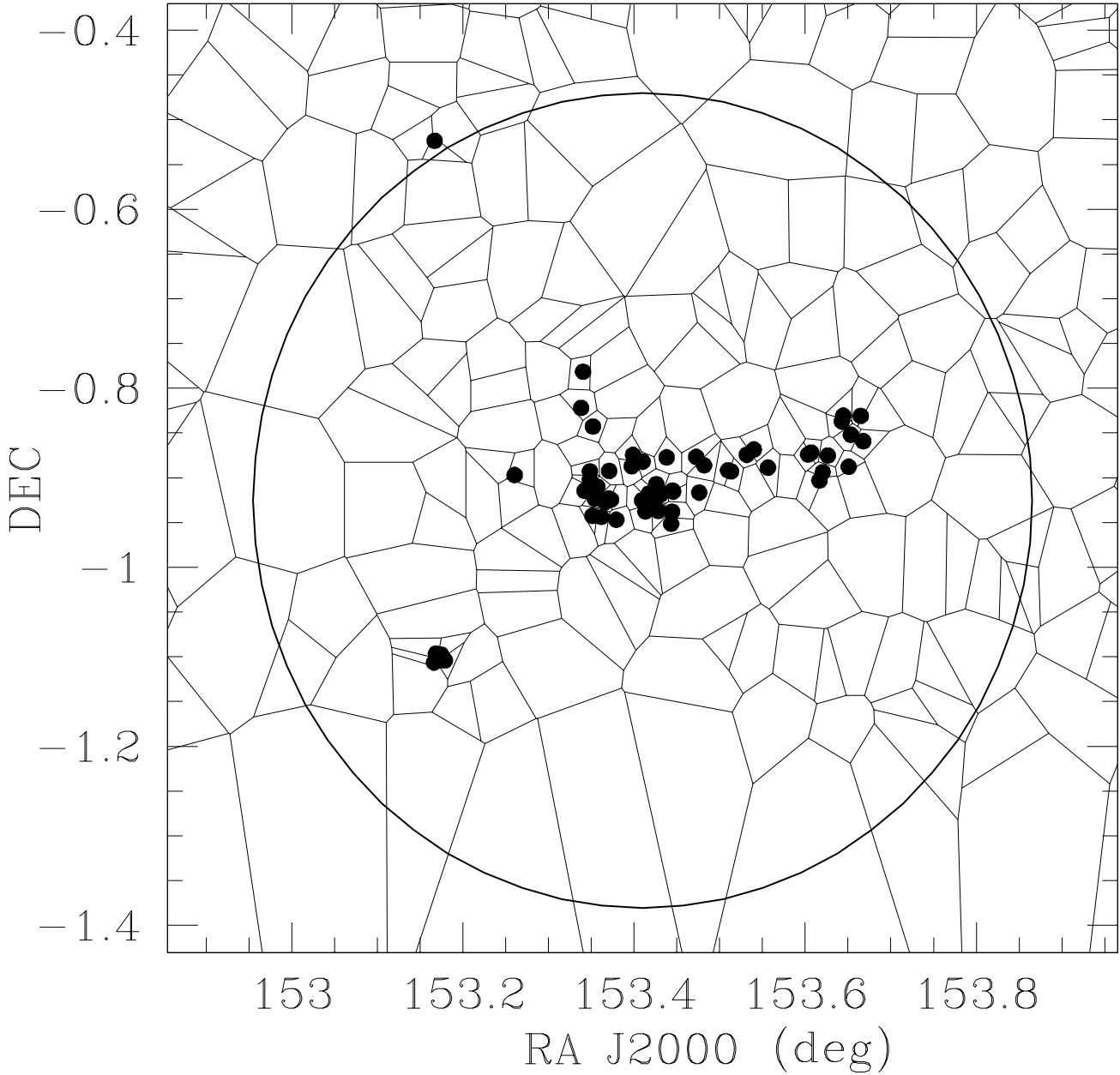}{5.8in}{0}{90}{90}{-290}{-140}
\caption{Same as Figure~3, but the Voronoi Tessellation is evaluated
only on the galaxies that satisfy the color-magnitude criteria used
in the VTT. See the solid lines in Figure~1(a) or Figure~2 for these
limits. Unlike Figure~\ref{fig:vttplot1}, the cluster is now
strikingly enhanced by the filled dots, which denote galaxies with $\delta
> 3$ evaluated from this Voronoi map.
\label{fig:vttplot2}}
\end{figure}

Once we have applied these cuts for a given redshift, we apply the
Voronoi Tessellation on the resulting distribution of galaxies.  We
then select all galaxies that satisfy $\delta > \delta_c$, where
$\delta_c$ is a constant overdensity threshold defined above.  We will set
this threshold to $\delta_c = 3$.
Figures~\ref{fig:vttplot1}~and~\ref{fig:vttplot2} illustrate this procedure:
Figure~\ref{fig:vttplot1} shows Voronoi Tessellation executed on all
galaxies in the region of Abell 957, whereas Figure~\ref{fig:vttplot2}
shows only those galaxies that satisfy the C-M cut for the cluster
redshift ($z=0.045$).  The large circle indicates a 1\mpc~ radius around
the cluster center.  In both cases, the dots highlight the overdense
galaxies with $\delta > 3$. Figure~\ref{fig:vttplot2} shows remarkable
enhancement of the cluster while Figure~\ref{fig:vttplot1} does not.  As
we can see, although the C-M limits are chosen generously in order to
ensure proper coverage of the observed cluster galaxy characteristics,
using these limits provides an important enhancement in the efficiency
of cluster selection due to the elimination of a significant
background population.

Once we have selected galaxies that highlight densely populated
regions, we have to identify regions with a concentration of these
``high-density'' galaxies.  This is done by selecting regions in which
the number of such galaxies, $N_{hdg}$, within a radius of 0.7\mpc~at
the assumed redshift, exceeds a certain threshold, such that $N_{hdg}
\geq N_{cut}$.  This is executed around each ``high-density'' galaxy,
and the center of a cluster is determined by the one that gives the
largest $N_{hdg}$.   We repeat this process as a function of redshift.
Once we obtain all the cluster candidates for each redshift bin, we go
through a process to filter out significant overlaps to finalize our
selection of clusters.  For every significant overlap we choose the
final cluster and its redshift to be the one that yields the largest
value of $N_{hdg}$. The distribution of $N_{hdg}$ with respect to
redshift is generally highly peaked and therefore justifies this
simple method of determining the redshifts.

Here, we take the high-density galaxy number cut $N_{cut}$ to be a
constant value, independent of redshift. This may introduce a bias
with redshift; the same cluster at higher redshift will contain fewer
galaxies, since the faint magnitude limit in the C-M cuts (5
magnitudes fainter than the BCG) soon exceeds the survey magnitude
limit as we go to higher redshift.  In order to find the optimal
threshold that changes with redshift, the algorithm needs to be tested
carefully to assure ourselves that the contamination level
(false-positives) does not increase too much.  We do not carry out
such an analysis here, as it requires quantitative assessment of
false-positives which is only possible with full N-body simulations,
with complete knowledge of cluster identities.  This would also
require proper assignments of colors for the background and cluster
galaxies.  Another way to investigate this matter is to use the real
data itself; although properly identifying false-positives can be a
slightly tricky business, we do address this issue of variable
$N_{cut}$ using visual inspection of cluster candidates in Paper II.
However for the current paper, we keep our VTT threshold $N_{cut}$
constant.  Therefore, as with every method, having such potential
biases motivates us to evaluate the selection function to assess the
fraction of clusters selected at each redshift and richness
(\S\ref{sec:sf}).  In addition, $N_{hdg}$ is not intended to measure
the richness of the cluster, but is rather a measure of the
significance of detection.  In Paper II, the final VTT selected
clusters will be run through the AMF fine filter for consistent
estimation of the cluster richness and redshift.

\section{Testing the Algorithms Using Simulated Clusters}
\label{sec:tests}
The MF was originally developed for the Palomar Distant Cluster Survey
(PDCS; P96) which is deep ($V \leq 23.8$) and narrow (5.1 \sqdeg), and
has been applied to similar data by others 
(e.g., Scodeggio \etal~1999 : 12 \sqdeg, $I
< 23$; Postman \etal~2001 : 16 \sqdeg, $I_{AB} < 24$).  Naturally, the
algorithm is optimized for this type of survey, whereas the SDSS is
shallower ($z < 0.5$) and much wider (Paper II will present results
for data over 150 \sqdeg, which is less than 2\% of the complete SDSS 
survey).  
The shallower depth makes the large-scale
structure variations much more pronounced in the two dimensional
distribution, which can affect the matched filter algorithm's
performance, as it assumes a uniform and homogeneous background
(BNP00).  Second, covering a large area increases the
probability of intersecting very nearby clusters ($z \lsim 0.05$)
which have angular extents as large as a few degrees. This increases
the rate of cluster overlaps, especially because we go to $z \sim 0.5$
for the SDSS.  In a 2D projection, discriminating between two
different overlapping clusters at two different redshifts using the
algorithms outlined above can be difficult unless the redshift gap is
sufficiently large (usually $\Delta z \gsim 0.4$);  this can affect the
completeness of intermediate redshift clusters.  The narrow pencil
beam surveys on the other hand, are carried out in regions known not
to have foreground clusters so they suffer less from this effect.
Wide angle, shallow surveys such as the APM survey or the EDSGC have
less range in redshift, and thus also suffer less from this effect.
Combining the local space density of Abell clusters and P96 results,
and assuming an unclustered distribution of clusters for simplicity
yields a $\sim 15\%$ rate of overlap for Abell richness class $\geq 0$
clusters (using a 1\mpc~radius for each cluster) for the redshift
range of $0.05 \leq z \leq 0.5$.  This rate of overlap will be further
enhanced by taking the non-zero two point correlation function into
account, and shall be addressed further in future work with the observed
space density of clusters from the SDSS.

Each method for detecting clusters has its own biases; moreover, the
sample selected depends sensitively on the detection threshold.  For
example, a 20\% change in $\sigma_{cut}$ can result in doubling the
final number of clusters in a given field.  Therefore it is
crucial for any cluster identification study to understand the
behavior of the results with respect to the selected thresholds.
Hence in this section we attempt to understand the parameters and
their limits that play a crucial role in determining the final cluster
selection, for the three algorithms.

Tests of the algorithms are performed on a set of artificial
clusters embedded in two different versions of a background galaxy
distribution: a uniform background and a clustered one. The latter is taken
to be the real galaxy distribution itself, from the SDSS.  The
major objectives for these tests are to determine the best
detection thresholds for the final cluster catalog, to
evaluate a realistic selection function for these thresholds
and to examine biases with respect to the local background density.

In the following we describe these tests, starting with a
brief description of the SDSS data that we used for the background. 

\subsection{Sloan Digital Sky Survey Imaging Data}
\label{sec:data}
The SDSS imaging data is taken with an imaging camera (Gunn
\etal~1998) on a wide-field 2.5 telescope, in five broad bands ($u, g,
r, i, z$) centered on [3551\AA, 4686\AA, 6166\AA, 7480\AA, and
8932\AA] respectively (Fukugita \etal~1996, Stoughton \etal~2001), in
{\it drift-scan} mode at sidereal rate. This results in an effective
exposure time of $54.1 s$, which yields a point source magnitude limit
of $r^* \approx 22.5$ (at $1.5''$ seeing).  More details of the
observations and data are covered in York \etal~(2000), Stoughton
\etal~(2001) and Paper II (and references therein).

The data used here and in Paper II are Equatorial scan data taken in
September 1998 during the early part of the SDSS commissioning phase,
and are part of the SDSS Early Data Release (runs 94 and 125;
Stoughton \etal~2001).  A contiguous area of about 150 \sqdeg\/ was
obtained during two nights, where the seeing varied from $1.1''$ to
$2.5''$ (85\% of the data was below $1.8''$), and coordinates ranging
from $-5^{\circ} < \mbox{RA} < 55^{\circ}$ and $-1.25^{\circ} <
\mbox{Dec} < 1.25^{\circ}$.  We include galaxies to $r^*=21^m$
(Petrosian magnitude; see below), a conservative limit at which
star-galaxy separation is reliable (see Paper II for details).  For
the present paper, a subset of 25 \sqdeg\/ from this data was taken
for the test region described in \S\ref{sec:montecarlo}.  The
coordinates of this region are $10^{\circ} < \mbox{RA} < 20^{\circ}$
and $-1.25^{\circ} < \mbox{Dec} < 1.25^{\circ}$, which was chosen
because it exhibits prominent large scale structure and clumpiness.

The magnitude of all objects quoted here are measured in Petrosian
quantities (Petrosian~1976) through the SDSS photometric pipeline
(Lupton \etal~2001). However, the colors of each object
quoted are computed from ``model magnitudes''.  Each galaxy is fit to
two profiles in the $r$ band : a de Vaucouleurs law and an
exponential law.  The model magnitudes in all five bands are computed
from the better of the two $r$ band fits; the colors obtained from
model magnitudes are thus a meaningful quantity as it uses the same
profile in all bands.

\subsection{The Uniform Background Case and the Hybrid Matched Filter}
\label{sec:uniformbg}
In order to produce a uniform galaxy background distribution, we took
the 25 \sqdeg\/ region from the SDSS data described above, and randomly
repositioned the galaxies while keeping their photometric
properties fixed. This creates a uniform galaxy background distribution
while ensuring that the galaxies otherwise have SDSS-like properties
(luminosity function and colors). The number-magnitude relation
$N(r^*)$ to the limiting magnitude $r^*= 21$ is presented in Yasuda
\etal~(2001): it shows power-law behaviour to $r^*=17$, curving below
this model at fainter magnitudes, as expected from K-corrections and
relativistic corrections.

Artificial clusters with 6 different richnesses were embedded at 8
different redshifts each, giving a total of 48 clusters with different
properties. The clusters were generated with a Schechter luminosity
function ($\alpha = -1.1, \, M_*(r^*) = -21.5 $; Blanton \etal~2001)
and a modified Plummer law radial profile ($r_{\mbox{max}} = 1
h^{-1}\mbox{Mpc}, \, r_{\mbox{core}} = 0.1 r_{\mbox{max}}$; see
Eq.~(\ref{eq:plummer})). Each cluster, properly normalized according
to its richness $\Lambda_{cl}$, was placed at the corresponding
redshift and then trimmed to the survey magnitude limit ($r^*=21$).
The ($g^*-r^*$) color for each galaxy was assigned characteristic for a
cluster at each redshift (as shown in Fig.~\ref{fig:cmlimits}).  The
insertion of simulated cluster galaxies increased the total number of
galaxies in the 25 \sqdeg\/ region by 7\%, to $105,600$ galaxies in total.  The
upper panel of Figure~\ref{fig:clusindata} shows the distribution of
the 48 artificial clusters over a $10 \times 2.5$ deg area (the
parameters of the clusters themselves are given in the figure
caption), and the resulting distribution when inserted into our
uniform background is shown in the middle panel.

All three cluster finding algorithms (MF; AMF; VTT) were first run on
this distribution of clusters in a uniform background of galaxies.
The goal for this was to test each algorithm in the simplest case, and
to find a reasonable detection threshold for each of them that
maximizes the number of successful detections, while keeping the false
detection rate minimal.  Although this uniform background case is far
from realistic, it has the advantage of unambiguously recognizing
false detections.  However, note that the number of clusters inserted
and their distribution of $z$ and $\Lambda_{cl}$ are arbitrary. Thus,
neither the fraction of false detections nor the absolute
value of the recovery fraction that we quote below have physical
significance; it is the relative values for different algorithms that
should be noted.

Figure~\ref{fig:detcurve} shows the detection efficiencies for all three
algorithms; each panel shows the
number of successfully recovered clusters (solid curve) and the number
of false detections (dotted curve) for one of the algorithms 
as a function of the detection
threshold for the 48 clusters inserted into the uniform background.
As the detection thresholds are decreased (going rightward in each
panel of Figure~\ref{fig:detcurve}), the number of successful
detections increases, but naturally the rate of false detections due
to Poisson statistics increases as well. Beyond a certain threshold
value, the number of false detections start to increase extremely rapidly,
while the number of successful detections only increases slowly; this
implies a large drop in efficiency. Figure~\ref{fig:detrate} shows this
in another way, plotting the rate at which the number of
successful detection increases as a function of the number of false
detections for all three algorithms.  We thus find a cut at which the
success rate starts to flatten out with respect to the false detection
rate; the vertical dotted line in Figure~\ref{fig:detrate} shows an
appropriate choice drawn just after the steepest part of the
efficiency curve.  This cut gives 14 false detections 
over 25 \sqdeg\/ for all three methods,
$0.56 \mbox{deg}^{-2}$, an acceptable level given the expected surface
density of real systems ($>5 \mbox{deg}^{-2}$; see Paper II).   The
vertical dashed lines in each panel of Figure~\ref{fig:detcurve} also
show this cut,
corresponding to $\sigma_{cut} = 5.5$ for the MF and ${\cal
L}_{cut} = 210$ for the AMF.  For the VTT the line indicates
$N_{cut} = 9$ (with a constant density contrast cut of $\delta_c =
3$).  We will refer to this cut in subsequent sections 
as the detection limit chosen for the uniform background case (note that
$\sigma_{cut}$ will be slightly lower for the final results; \S 4.1).
As Figure~\ref{fig:detrate} demonstrates, of all methods, the
MF is the most efficient in recovering clusters for a given number of
false detections.

The additional clusters detected by the MF are clusters with a weaker
signal (low $\Lambda_{cl}$, high $z$), as seen in
Figures~\ref{fig:unif_results} a~and~b.  These results are obtained
using the cuts determined above.  Although the AMF (crosses in
Figure~\ref{fig:unif_results}a) should in principle converge to the MF
(squares in Figure~\ref{fig:unif_results}a) in the 2-D case, they
differ for weak clusters due to the fact that
their peak selection method in the final step is done differently
(refer to \S\ref{sec:mf}; Table~\ref{tab:mf}).  The parameter
evaluation, on the other hand, is most accurate with the AMF fine
filter.  Figures~\ref{fig:unif_results} c~and~d show the input
parameters for the clusters, $z$ and $\Lambda_{cl}$, plotted against
the parameters evaluated by the AMF fine filter. They include the
additional clusters found by the MF.  The MF $z$ estimates show
a factor of $\sim 1.5$ increase in standard deviation for higher redshift
clusters ($z > 0.25$), which translates into a larger deviation in
$\Lambda_{cl}$ as well, by a similar amount.  This result thus calls for a
marriage of the two algorithms in order to maximize the efficiency of
the final result.  We have therefore adopted the following hybrid
method for the final Matched Filter based cluster finder (HMF: Hybrid
Matched Filter), \\ \\
\noindent
\hskip 0.5in $\bullet$ First the MF creates likelihood maps. \\
\hskip 0.5in $\bullet$ We threshold at $\sigma_{det} = \sigma_{cut}$ according to the MF to choose final cluster candidates. \\
\hskip 0.5in $\bullet$ Finally, we evaluate the AMF fine filter on these cluster positions to determine $\Lambda_{cl}$ and $z_{est}$. \\
\\
We use this recipe as the standard Matched Filter method
throughout the rest of this paper.

\subsection{Monte Carlo with a ``Realistic'' Background}
\label{sec:montecarlo}
\begin{figure}[p]
\plotfiddle{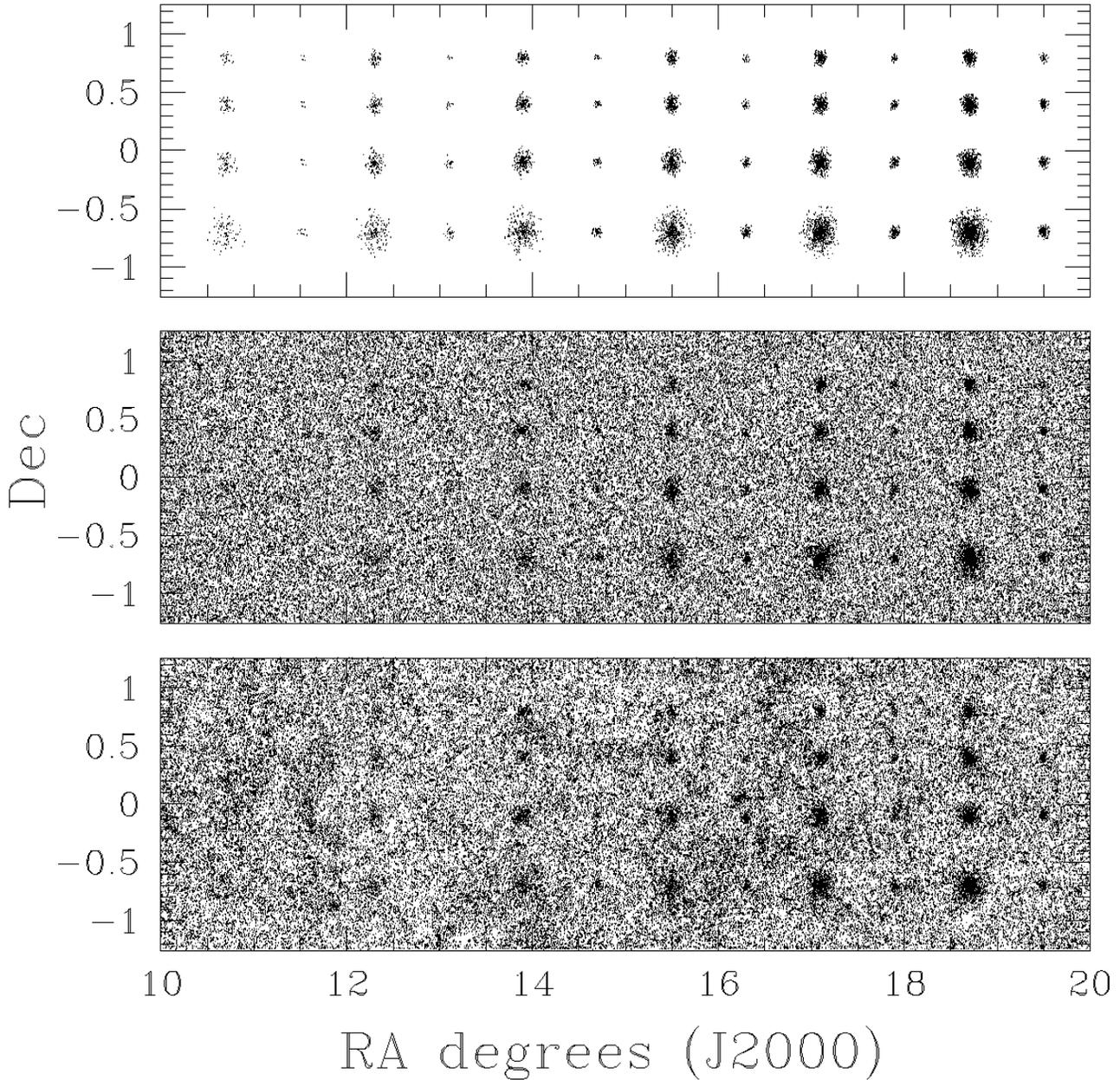}{5.8in}{0}{90}{90}{-290}{-50}	
\caption{Distribution of simulated clusters with different
richnesses and redshifts is shown in the top panel. The clusters are
distributed such that the richness increases from left to right, and
the redshift from bottom to top.  The values are $\Lambda_{cl} = [ 20,
40, 70, 110, 160, 220 ], \, z = [ 0.08, 0.14, 0.20, 0.26, 0.32, 0.38,
0.44, 0.5 ]$. Every two columns correspond to a cluster of a
given richness at eight different redshifts.  The bottom panel
shows these clusters embedded in a 25 \sqdeg~region of SDSS
equatorial scan data. The middle panel shows the same clusters embedded in
a uniform background generated by randomly repositioning the
background galaxies shown in the bottom.
\label{fig:clusindata}}
\end{figure}

The previous test, using a uniform background, allowed us to determine
the detection threshold necessary to minimize false detections due to
projections in random fields. We need to apply a realistic background
in order to get an accurate determination of the selection function,
including the effect of large-scale structure and cluster projection
along the line-of-sight.  In relatively shallow surveys like the APM
or the EDSGC the cluster detection efficiency has been shown to depend
significantly on the local background density (see BNP00).
The SDSS is considerably deeper ($z \sim 0.5$ compared to $z \sim
0.2$) and therefore this effect should be less dramatic, due to projection 
in 2D of the real large-scale structure.  Nevertheless as
the lower panel of Figure~\ref{fig:clusindata} shows, the SDSS galaxy
distribution is still far from uniform, and the effect of large-scale
structure on the selection needs to be quantified.

We now place the same set of simulated clusters on the real SDSS
distribution as shown in the lower panel of Figure~\ref{fig:clusindata}.
The effect of the real background is immediately noticeable even by
eye; the low richness clusters ($\Lambda_{cl} \leq 40$) are nearly
washed out and the intermediate richness clusters
($\Lambda_{cl} = 70, 110$) start to blend in with the clumpy
background.

The 48 simulated clusters with the properties given in
Figure~\ref{fig:clusindata} were inserted into the data at random
positions (unlike the grid distribution in the lower panel of
Figure~\ref{fig:clusindata}), but avoiding overlap with one
another. Both the HMF and the VTT were run on this catalog. This was
repeated 100 times.  We use the results to test 
both the detection and recovery of clusters, and
the evaluation of their parameters, $z$ and $\Lambda_{cl}$.  In the
next section we discuss the results from these tests in three
parts. First, we study the effect of the imposed detection
limits; second, we evaluate the selection function; and finally we
examine the dependence of the detection efficiency and the recovered
parameters on the local background.

\section{Results}
\subsection{Detection Limits}
\label{sec:det_limit}
A set of clusters embedded in a uniform background distribution is an
ideal case.  In the real universe, non-uniformity comes in many forms:
from large scale modulations (great wall, voids etc.) down to small
scale fluctuations (e.g., compact groups, close pairs and triplets
etc.), in addition to actual clusters of galaxies.  It is mainly the
small scale features aided by projection that will cause 
the false-positive rate to increase.  
Therefore, the false detection rate that was
determined with the uniform background ($0.56/{\mbox{deg}}^2$ for
$\sigma_{cut} = 5.5$ and $N_{cut} = 9$) can be regarded as a {\it
lower limit} for those thresholds in a realistic background.  In
addition, the structure exhibited on a range of scales diminishes the
contrast of real clusters, making them more difficult to recover.
This implies that these detection thresholds are {\it upper
limits} on what they should be in the real universe. As we will see below,
the detection threshold has to be set lower with a realistic
background in order to recover a similar range of clusters.

\begin{figure}[p]
\plotfiddle{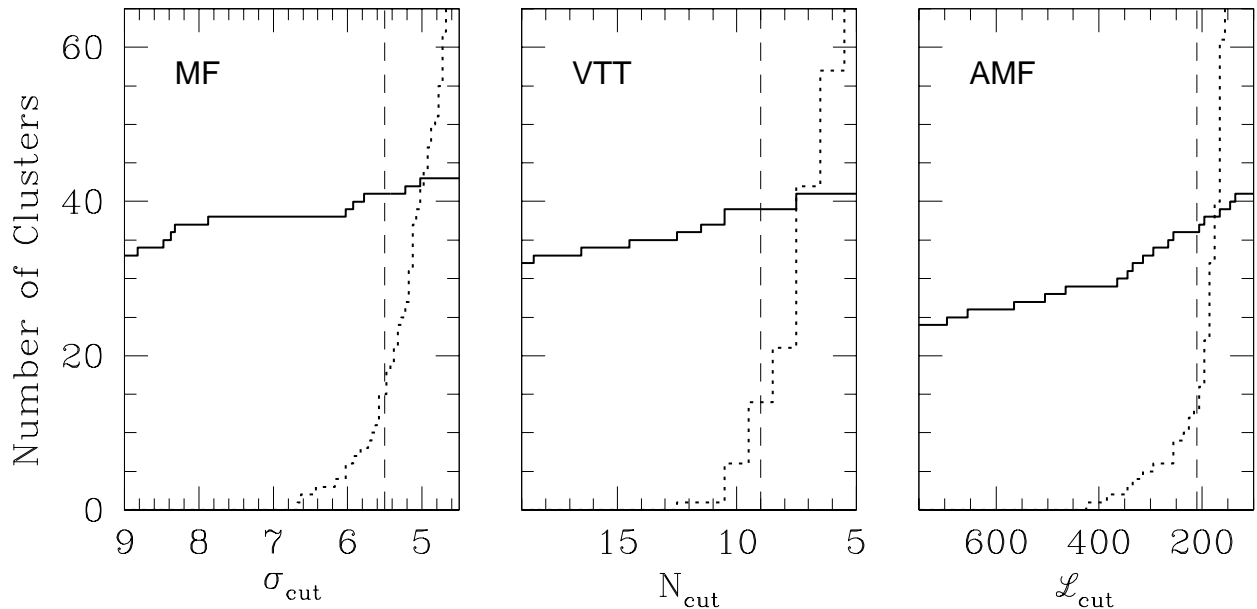}{5.8in}{0}{90}{90}{-290}{-340}
\caption{The number of successful detections (solid) and the number of
false detections (dotted) as a function of the detection threshold for
each cluster finding algorithm: MF (left), VTT (middle), and AMF
(right).  These results are from 48 clusters inserted into a uniform
background (see middle panel of Fig.~\ref{fig:clusindata}).  The
MF is more efficient in detecting clusters than the AMF, due to
differences in the thresholding method.  The vertical dashed line is
drawn at the thresholds that yield 14 false detections in a uniform
background of 25 \sqdeg; this corresponds to $\sigma_{cut}= 5.5$,
$N_{cut} = 9$ and ${\cal L}_{cut} = 210$, yielding maximum
completeness while still keeping the false detection rate less than
10\% of the expected surface density of real clusters.  Note that for
all three algorithms, the ranges shown for the detection thresholds 
are calibrated to yield similar numbers of false detections.
\label{fig:detcurve}}

\end{figure}

\begin{figure}[p]
\plotfiddle{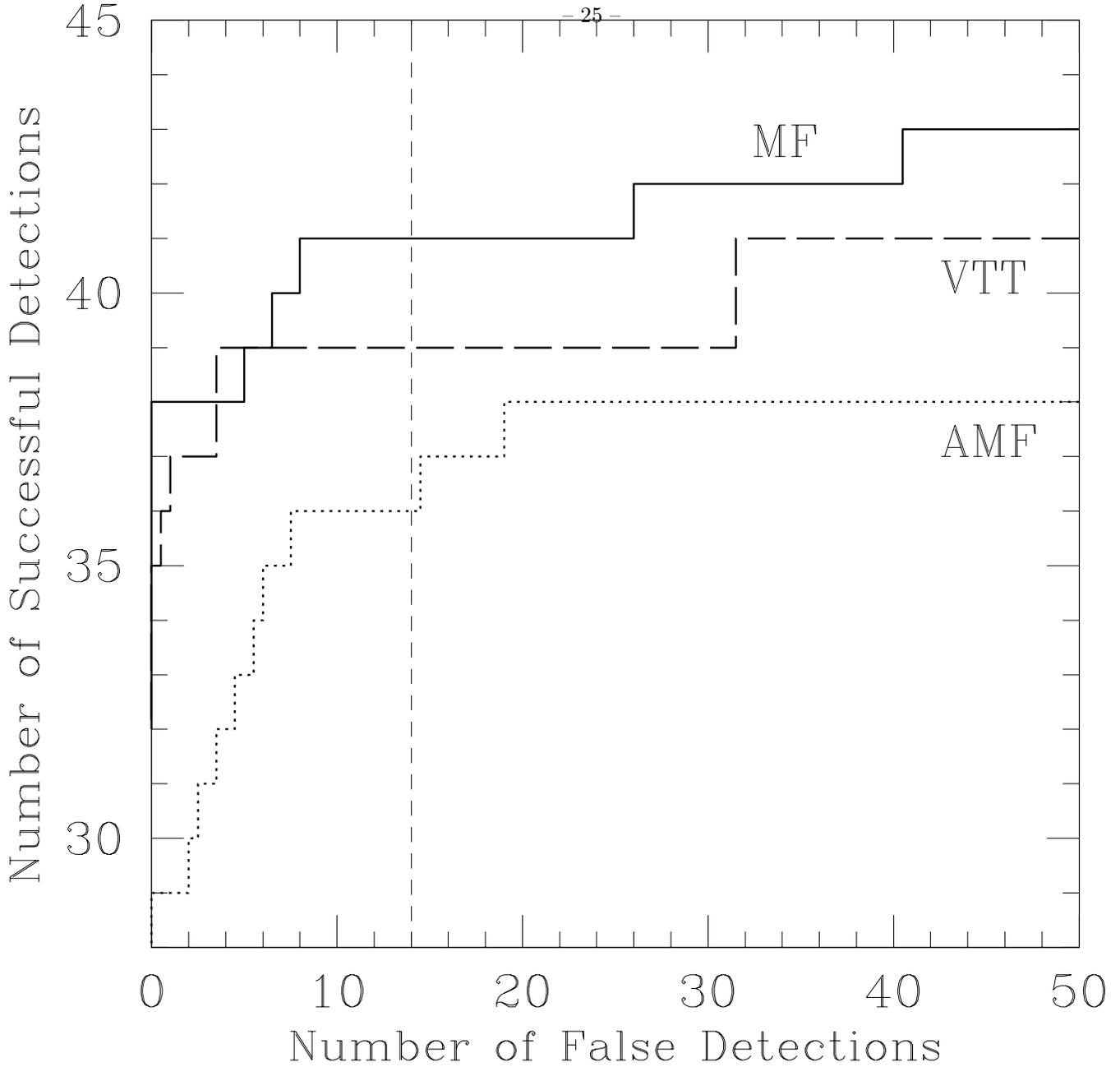}{5.8in}{0}{90}{90}{-290}{-140} 
\caption{ The
number of successful detections as a function of the number of false
detections, as the detection threshold is changed for all three
algorithms, in the uniform background case.  This is an alternative 
representation of the data in Figure~\ref{fig:detcurve}.  The ranges for the 
detection thresholds represented here are 
$7 \geq \sigma_{cut} \geq 4.5$ for the HMF
(solid), $13 \geq N_{cut} \geq 5$ for the VTT (dashed) and $450 \geq
{\cal L}_{cut} \geq 160$ for the AMF (dotted). The vertical dotted line
shows the cut for $\sigma_{cut}= 5.5$, $N_{cut} = 9$ and 
${\cal L}_{cut} = 210$, at which the rapid increase of
success rate stops with respect to the number of false detections (same cut 
as in Fig.~\ref{fig:detcurve}).
For a given number of false detections, the MF is most efficient in recovering
clusters in a uniform background.
\label{fig:detrate}}
\end{figure}

\begin{figure}[p]
\plotfiddle{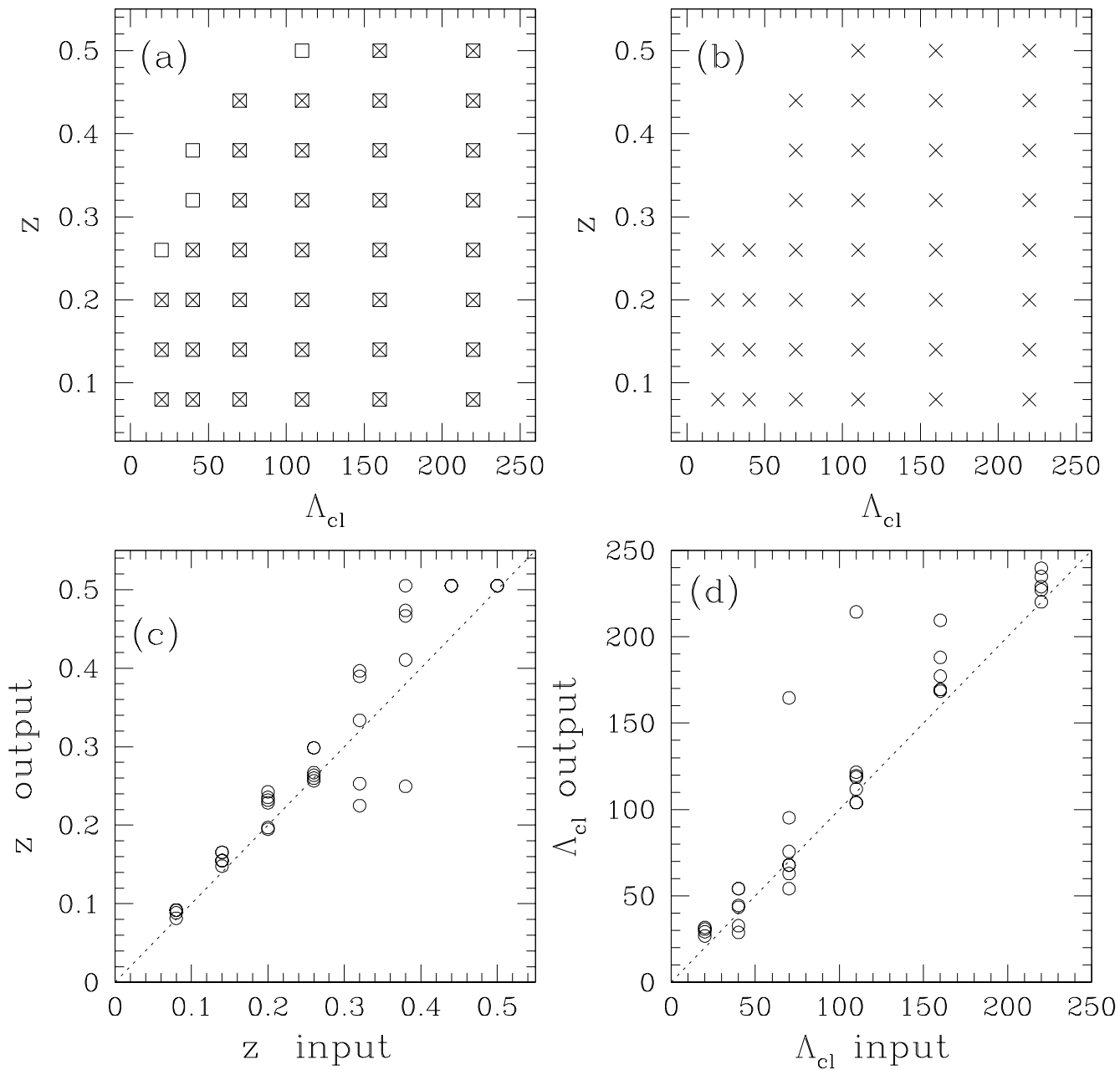}{5.8in}{0}{90}{90}{-290}{-140}
\caption{ The results for 48 clusters embedded in a uniform
background, using the detection threshold from
Figure~\ref{fig:detcurve} (see caption), 
which yields 14 false detections for each techniques.  (a)
The parameters of the recovered clusters by the MF (squares) and the
AMF (crosses).  The MF selection method is more sensitive to weaker
signals, i.e., higher $z$ and lower richness. (b) The clusters that were
recovered by the VTT. The VTT is also slightly less efficient than the MF but
more efficient than the AMF. 
All three algorithms show similar recovery patterns. 
(c) AMF fine filter
determination of redshift for MF selected clusters (squares in (a)), 
versus the input value. (d)
Same as (c), for the richness measure, $\Lambda_{cl}$. (c) and (d) are
equivalent to the results for the HMF (Hybrid Matched Filter; see text
for details).
\label{fig:unif_results}}
\end{figure}

\begin{figure}[p]
\plotfiddle{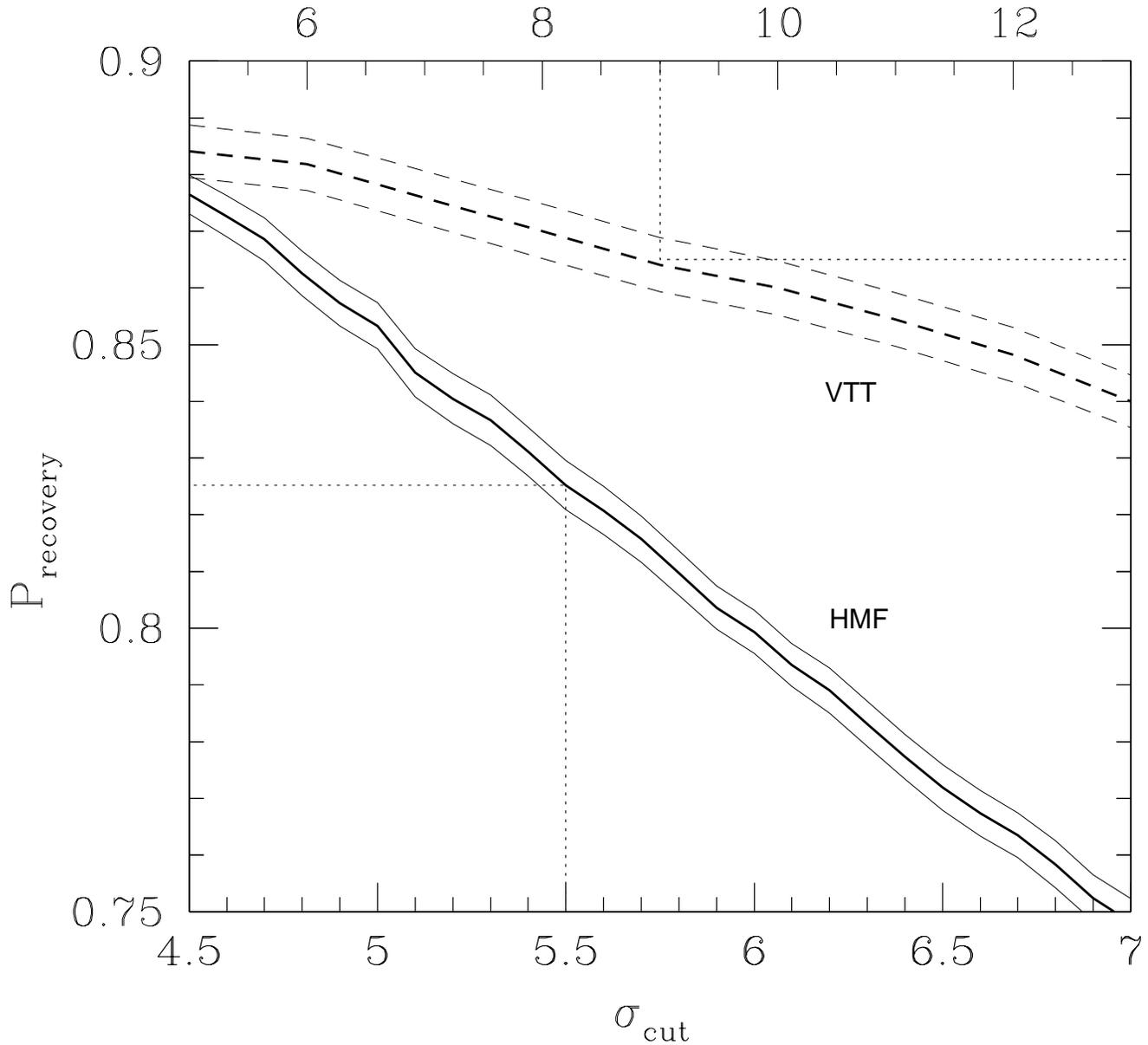}{5.8in}{0}{90}{90}{-290}{-140}
\caption{ The recovery rate of clusters in our Monte Carlo experiment
as a function of the detection thresholds $\sigma_{cut}$,
$N_{cut}$ for the HMF (solid) and the VTT (dashed)
respectively. These recovery rates are evaluated relative to the
clusters that were detected in the uniform background case, i.e., the
average fraction of those clusters recovered in 100 Monte Carlo
realizations.  The standard deviations between realizations are traced
with thin lines. The range of $\sigma_{cut}$ and $N_{cut}$
shown here are coincident with the values that were used to
plot Figure~\ref{fig:detrate}. The dotted lines show the performance
for $\sigma_{cut} = 5.5$ and $N_{cut} = 9$,
which are the values that were determined from the uniform background
case (vertical lines in Fig.~\ref{fig:detcurve}~and~\ref{fig:detrate}).
\label{fig:recovery_ratio}
}
\end{figure}

For each cluster finder, we determine how many of the clusters that
were detected in the uniform background case (with the corresponding
detection thresholds $\sigma_{cut} = 5.5$ and $N_{c} = 9$; see
Figure~\ref{fig:unif_results}), were also recovered in each Monte
Carlo realization with the SDSS galaxy background.  This recovery
ratio was evaluated for different detection thresholds, and averaged
over all realizations.  Figure~\ref{fig:recovery_ratio} shows the
results for both the HMF and the VTT, for a range of detection
thresholds that are equivalent to those used in
Figure~\ref{fig:detrate}.  This recovery ratio (relative to the 
uniform background case) changes quite rapidly
as a function of $\sigma_{cut}$ for the HMF, but stays rather robust
as a function of $N_{cut}$ for the VTT.  This comparison can be misleading
since there is no formal correspondence between $N_{cut}$ and
$\sigma_{cut}$.  However, as noted above, these ranges are calibrated to yield
approximately the same number of false detections in a uniform
background distribution (c.f., Fig.~\ref{fig:detcurve}).  
Given this relation as a yardstick, the
noticeable difference in the slope of the two curves suggests that the
Matched Filter algorithm results will be more sensitive to the
detection threshold that is chosen, while a sample based on 
the VTT is robust to the exact value of $N_{cut}$.  This implies
in addition, that the Matched Filter is more subject to the effects of
a non-uniform background, i.e., the detection efficiency is
affected by the background distribution.  

This argument is further supported by the lower recovery rate of the
HMF shown in Figure~\ref{fig:recovery_ratio} at the thresholds of
$\sigma_{cut} = 5.5$ for the HMF and $N_{cut} = 9$ for the VTT (dotted
lines; where both algorithms yielded 14 false detections in the
uniform background case).  Thus the detection efficiency of the HMF
appears to be more affected by a non-uniform background than is the
VTT.  Indeed, this is not too surprising as the Matched Filter
algorithm explicitly assumes a uniform background in its model.  In
order for the HMF to achieve the same recovery rate as the VTT, the
detection threshold would need to be lowered to a value of
$\sigma_{cut} \sim 4.7$ (which greatly increases the number of false
detections, this would yield $\sim 60$ false detections in a uniform
background; Fig.~\ref{fig:detcurve}).  These fractional recovery rates
refer to the sample of 48 clusters over their entire range of $z$ and
$\Lambda_{cl}$; hence it does not apply to an observed sample of
clusters with a true richness function distribution.  The clusters that
are missed in the selection are those at the weaker end of the
distribution of signal, which are the most abundant in the universe
(i.e., poor, distant), therefore the gap between the recovery fractions
shown above could be even larger for a more realistic
configuration.

\begin{figure}[p]
\plotfiddle{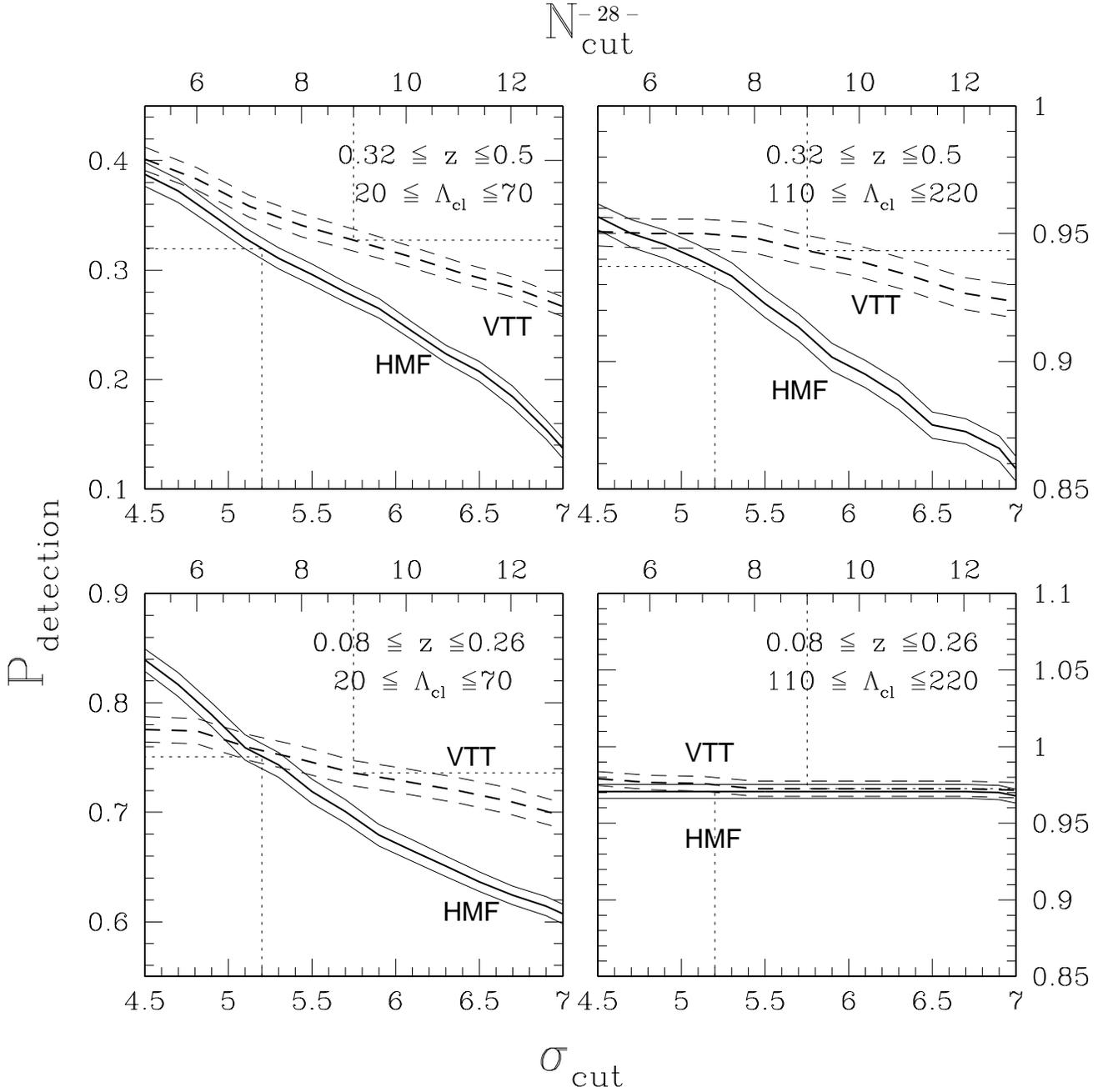}{5.8in}{0}{90}{90}{-290}{-140}
\caption{Similar to Figure~\ref{fig:recovery_ratio}, but this shows
the absolute recovery rates of clusters in 4 different ranges of
cluster parameters for the HMF (solid) and the VTT (dashed), averaged
over 100 Monte Carlo realizations. The $1 \sigma$ dispersion is traced
with thin curves.  Four different panels show those for poor low $z$
clusters (lower left), poor high $z$ clusters (upper left), rich low
$z$ clusters (lower right), and rich high $z$ clusters (upper
right). Both algorithms agree very well for clusters with the highest
signals (rich, low $z$) but VTT does slightly better in general for the
thresholds determined from the uniform background case:
$\sigma_{cut}=5.5$ and $N_{cut}=9$.  The dotted lines show where
$N_{cut}=9$ for the VTT, and $\sigma_{cut}=5.2$ for the HMF, lowered
to this value to match the performance of the VTT.
\label{fig:detection_ratio}
}
\end{figure}

These result do not immediately mean, however, that the VTT does better
overall.  Recall from Figure~\ref{fig:detcurve} that the HMF was more
efficient in recovering clusters in the uniform background to begin
with.  Given the right choice of detection thresholds we show that
their performances are in fact similar in the end.  
Figure~\ref{fig:detection_ratio} shows the
{\it absolute} recovery fraction of inserted clusters, rather than that
{\it relative} to the uniform background (Fig.~\ref{fig:recovery_ratio}),
for both the HMF (solid) and the VTT (dashed).  
This is shown in four
different subgroups in cluster parameter space: low $z$, low
$\Lambda_{cl}$ (lower left); high $z$, low $\Lambda_{cl}$ (upper left);
low $z$, high $\Lambda_{cl}$ (lower right); and high $z$, high
$\Lambda_{cl}$ (upper right).  For clusters with the strongest signals
(rich and nearby), both algorithms agree well with very high
efficiencies, constant with respect to the detection thresholds.  For
clusters with weakest signals (poor, high $z$) both methods have very
low recovery rate.  Apart from the rich nearby regime, the HMF indeed
shows slightly lower efficiencies than the VTT for the thresholds
determined above ($\sigma_{cut} = 5.5$, $N_{cut} = 9$) suggesting a drop
in efficiency due to the presence of a non-uniform
background. However, the slopes of the HMF efficiency curves are much
steeper, and allow us to bring up the detection efficiency easily to
the VTT performance level, by lowering the detection threshold
slightly to $\sigma_{cut} = 5.2$.
This naturally increases the false detection rate of the HMF 
in the uniform background case (refer to
Fig.~\ref{fig:detcurve}) by approximately 80\% to $1 \,\mbox{deg}^{-2}$, 
but as it is less than
20\% of the expected surface density of real clusters (Paper II) we
adopt this new value for $\sigma_{cut}$ as appropriate for
selecting clusters from the SDSS imaging data. The final detection
thresholds determined for both algorithms, $\sigma_{cut}=5.2$ and
$N_{cut}=9$, are shown as dotted lines in Figure~\ref{fig:detection_ratio},
indicating that the performance of the two algorithms is very similar.

\subsection{Cluster Selection Function}
\label{sec:sf}
\begin{figure}[p]
\plotfiddle{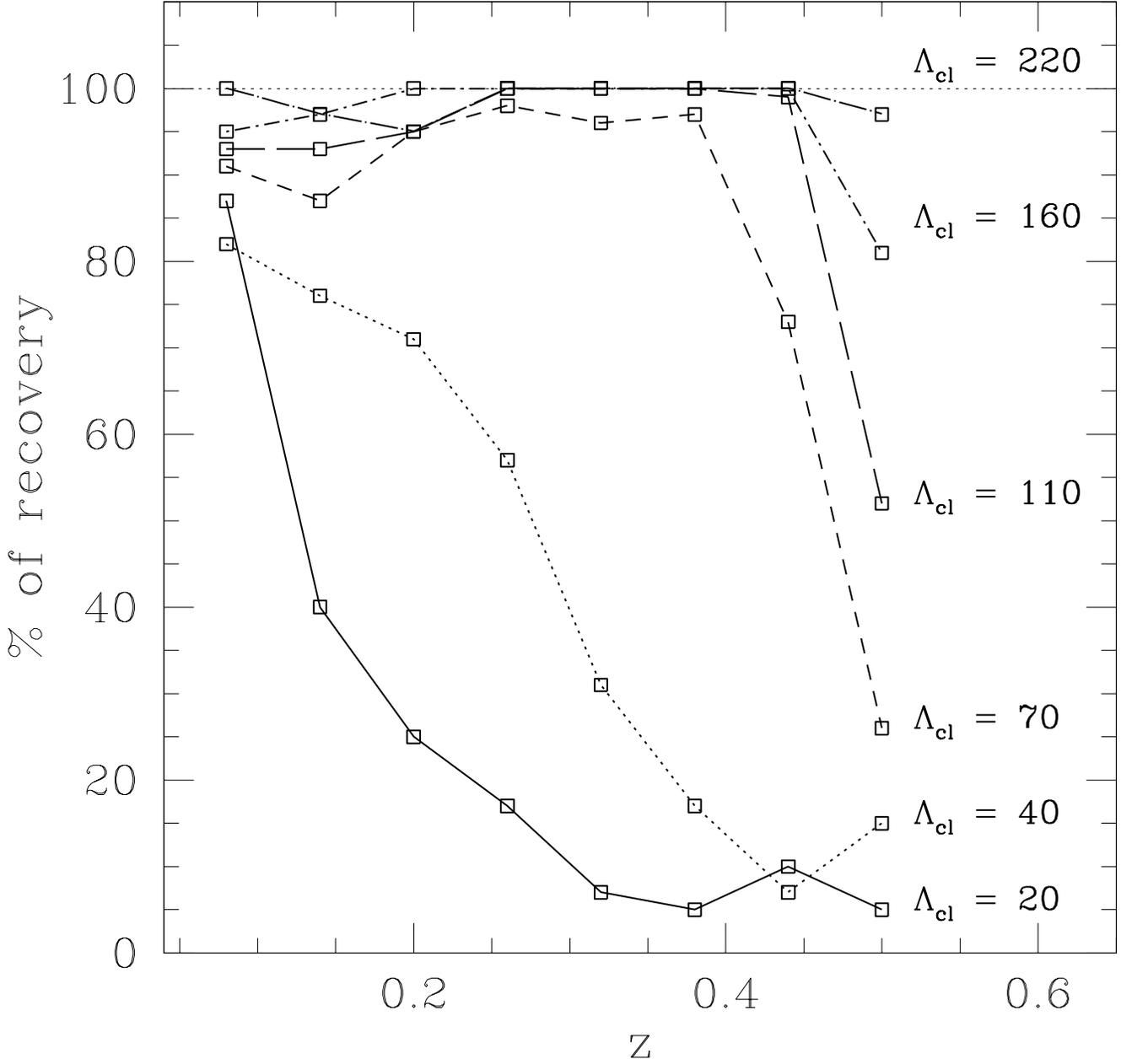}{5.8in}{0}{90}{90}{-290}{-140}
\caption{ The selection function evaluated for the HMF
($\sigma_{cut}=5.2$) as a function of redshift for clusters with
different richness measures $\Lambda_{cl}$.   Both $z$ and
$\Lambda_{cl}$ shown are the input values of the simulated clusters
generated with a Schechter luminosity function using $M_r^* = -21.7$
and $\alpha = -1.1$, and a modified Plummer law profile with $r_{max}
= 1$\mpc~and $r_{c} = 0.1$\mpc.
\label{fig:mfSF}}
\end{figure}

\begin{figure}[p]
\plotfiddle{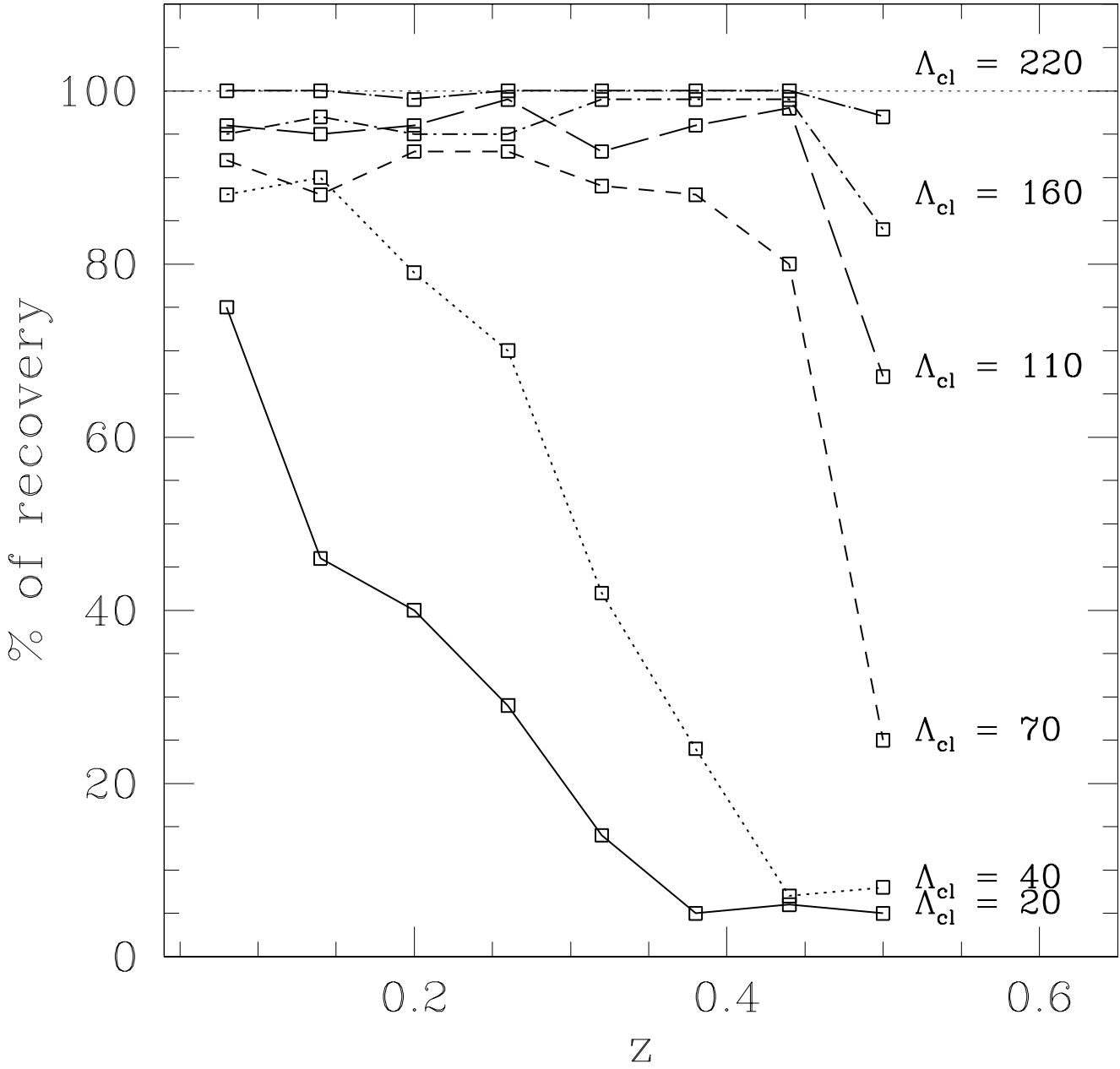}{5.8in}{0}{90}{90}{-290}{-140}
\caption{ The selection function evaluated for the VTT ($N_{cut}=9$),
in the same Monte Carlo experiment as Figure~\ref{fig:mfSF}.
\label{fig:vttSF}}
\end{figure}

With these detection thresholds, we now assess the overall
performances of both cluster finders for these values in more detail,
an important task for any study that requires a complete sample.  We
present selection functions evaluated from the fraction of clusters in
each redshift and richness class that are recovered in the Monte Carlo
simulations (using a realistic background).  These selection functions
for the HMF and the VTT are shown in
Figures~\ref{fig:mfSF}~and~\ref{fig:vttSF} with $\sigma_{cut}=5.2$ and
$N_{cut} = 9$, respectively.  The two methods are qualitatively
similar in terms of their overall performances; for clusters with
$\Lambda_{cl} \geq 70$, over 80\% are recovered to redshift $z \sim
0.45$, falling off rapidly as one goes to $z \sim 0.5$.  There are
still slight differences; the HMF does better in the high
$\Lambda_{cl}$ intermediate $z$ ($0.2 \lsim z \lsim 0.4$) domain,
while the VTT seems to perform slightly better for clusters with low
$\Lambda_{cl}$ and low $z$ ($z \lsim 0.2$).  This can be explained in
terms of the generous C-M cuts adopted for the VTT, intended to
account for possible fluctuations in the cluster color-magnitude
properties.  As a result, as one goes to higher redshift, where the
C-M limits start to move in towards the core of the C-M distribution
of normal galaxies (see Fig.~\ref{fig:cmlimits}), the population of
galaxies that are rejected by this procedure reduces significantly,
making this filtering less effective and therefore reducing the
efficiency of recovering the clusters.  At low redshifts $z \lsim
0.2$, on the other hand, it rejects most of the galaxy background,
making the search most efficient.  If we were to narrow these C-M
limits, we might bias ourselves against clusters with unusual
properties.  This will be tested with real SDSS clusters by
investigating their color-magnitude properties, to determine how
tightly we can impose the limits without biasing the cluster
selection.

We compare our selection functions to those of BNP00 (see their
Fig.~2), evaluated for cluster detections in the EDSGC data.  The
redshift range probed with the EDSGC is much shallower ($z \lsim
0.15$; $b_j < 20.5$, which corresponds roughly to $r < 19$ for a
typical elliptical galaxy) than that of the SDSS, however, they use
the same Monte Carlo technique by inserting simulated clusters in the
real data itself, making the comparison meaningful.  Recall that the
cluster detection completeness depends significantly on the use of
different background distributions, as we have shown in
Figures~\ref{fig:recovery_ratio}~and~\ref{fig:detection_ratio}.


Figure~2 of BNP00 shows the selection function for four different
richnesses as a function of redshift. Their richnesses correspond to
$\Lambda_{cl}=10.15, 20.3, 40.6, \mbox{and}~81.2$, allowing a
straightforward comparison to
Figures~\ref{fig:mfSF}~and~\ref{fig:vttSF}.  For $\Lambda_{cl}=20$
clusters, their efficiency drops from 70\% to 40\% over the redshift
range of $0.05\leq z \leq 0.15$.  Neither the HMF nor the VTT does
significantly better for the same range of parameters, and beyond $z
\sim 0.2$, clusters with $\Lambda_{cl}=20$ are virtually invisible
even in the SDSS.  However, for richer clusters, the SDSS clearly does
better. For $\Lambda_{cl} = 40$ clusters, BNP00 shows a steep drop in
efficiency to $\sim 60\%$ at $z = 0.15$, whereas the SDSS efficiencies
stay at $\sim 80\%$ out to $z=0.2$.  Similarly, the recovery of
$\Lambda_{cl} = 70$ clusters in the SDSS stays highly efficient
($\gsim 90\%$) and constant out to $z\sim 0.4$, while BNP00 efficiency
for $\Lambda_{cl} = 80$ clusters shows a clear drop from 100\% to 80\%
already at $z=0.15$.  The above comparison conforms to what we would
expect: While going deeper in magnitude helps the recovery of clusters
at higher redshifts, there is a limit in the cluster richness for
which this is true. In other words, poor clusters are hard to find at
high redshifts no matter what the depth of the survey is.  Apart from
those differences in the SDSS and EDSGC, the overall performances are
qualitatively similar, which is reassuring given that our HMF is quite
similar to their method.

\subsection{Dependence on Local Background}
\label{sec:bg}
\begin{figure}[p]
\plotfiddle{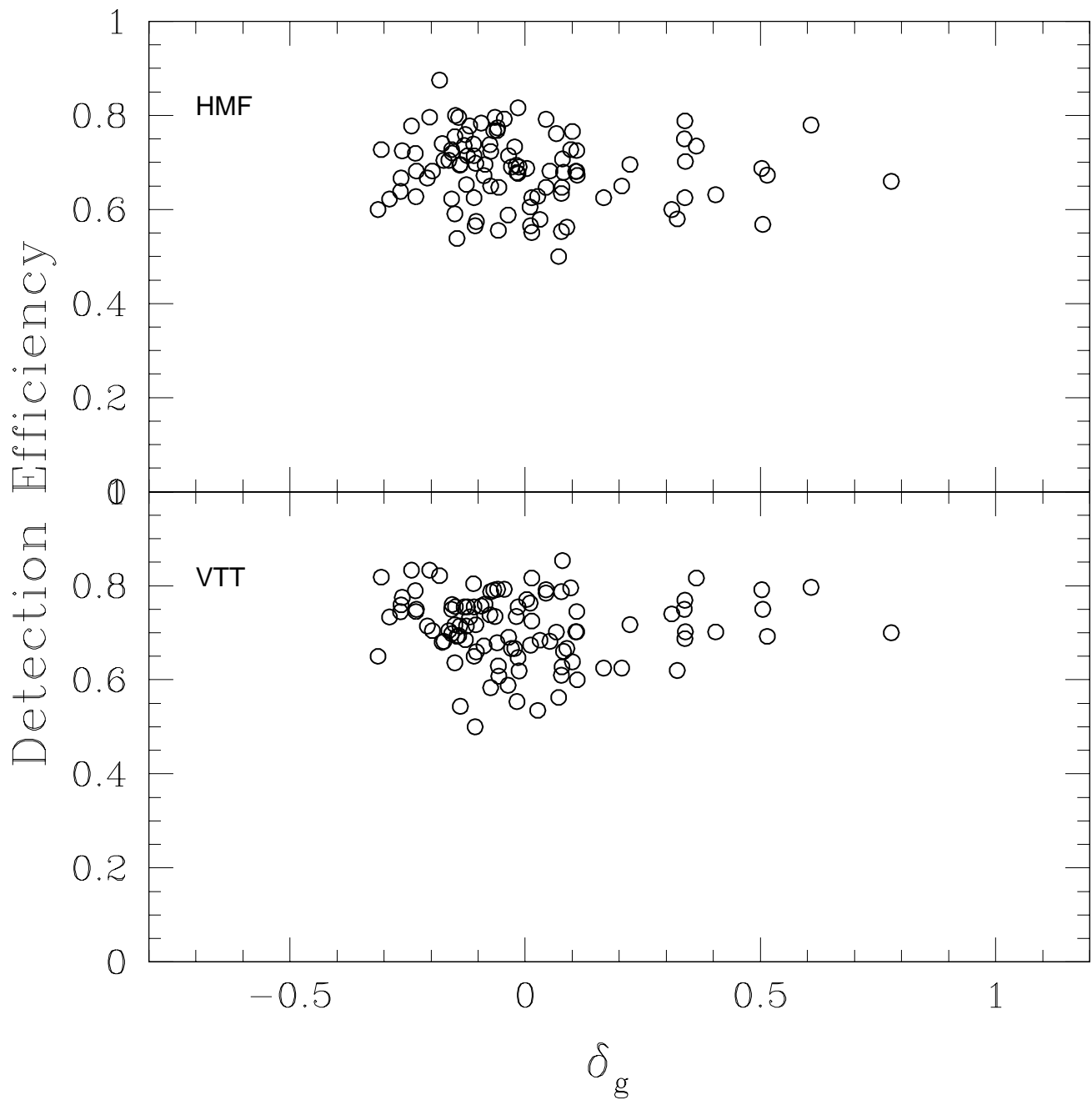}{5.8in}{0}{90}{90}{-290}{-140}
\caption{The detection efficiency as a function of the local
background density contrast. Each circle represents the average
fraction of clusters that were detected amongst those that were
inserted in each independent cell of area $(0.5\, \mbox{deg})^2$.  The
density contrast is $\delta_g = (N_g - \bar{N})/\bar{N}$, where $N_g$
is the number of background galaxies in each cell and $\bar{N}$ is the
average number of background galaxies in all cells. The top panel
shows the results for the HMF and the bottom for the VTT. The Spearman
Rank-Order Correlation test confirms that neither distribution shows any
correlation with the background density (see text).
\label{fig:cc_det}}
\end{figure}

The final issue we examine from this Monte Carlo experiment is the
extent to which the results depend on the local background density.
First, we look at the behavior of the detection efficiency at
different locations.  We divide our $2.5 \times 10$ degree Monte Carlo
region into $5 \times 20$ subregions, $\sim 0.5$ deg on a side.  This
bin size is chosen to be large enough to get an appreciable number of
clusters in each bin, for reliable statistics.  We then count the
number of background galaxies within each bin $N_g$ (excluding cluster
galaxies), and evaluate the detection efficiency of clusters that fall
into each bin.  Figure~\ref{fig:cc_det} shows a scatter plot of the
cluster detection efficiency as a function of the background density
contrast $\delta_g = (N_g - \bar{N})/\bar{N}$.  We have
performed Spearman Rank Correlation tests (Press \etal~1990) and find
that $r_s = -0.17, t=-1.72$ for the HMF and $r_s = -0.14, t=-1.39$ for
the VTT, where $r_s$ is the Spearman rank-order correlation
coefficient, and $t = r_s \sqrt{(N-2)/(1-r_s^2)}$ which is
distributed like a Student's t distribution with $N-2$ degrees of
freedom. This translates into correlations at only 5\% and 8\%
confidence for the HMF and VTT respectively, confirming that neither
method shows any significant correlation between their detection
efficiency and the local background density.
This is very encouraging,
considering the unambiguous dependence that BNP00 have
shown for their shallower sample (see Figure 3 in their paper).
This suggests that our SDSS sample is deep enough to subdue the
background fluctuations by projection effects, to a level that it does
not affect the detection efficiency significantly.  However, we have
seen in \S\ref{sec:det_limit} that the overall performance is
still highly influenced by the {\it presence} of a non-uniform
background, especially for the Matched Filter.

\begin{figure}[p]
\plotfiddle{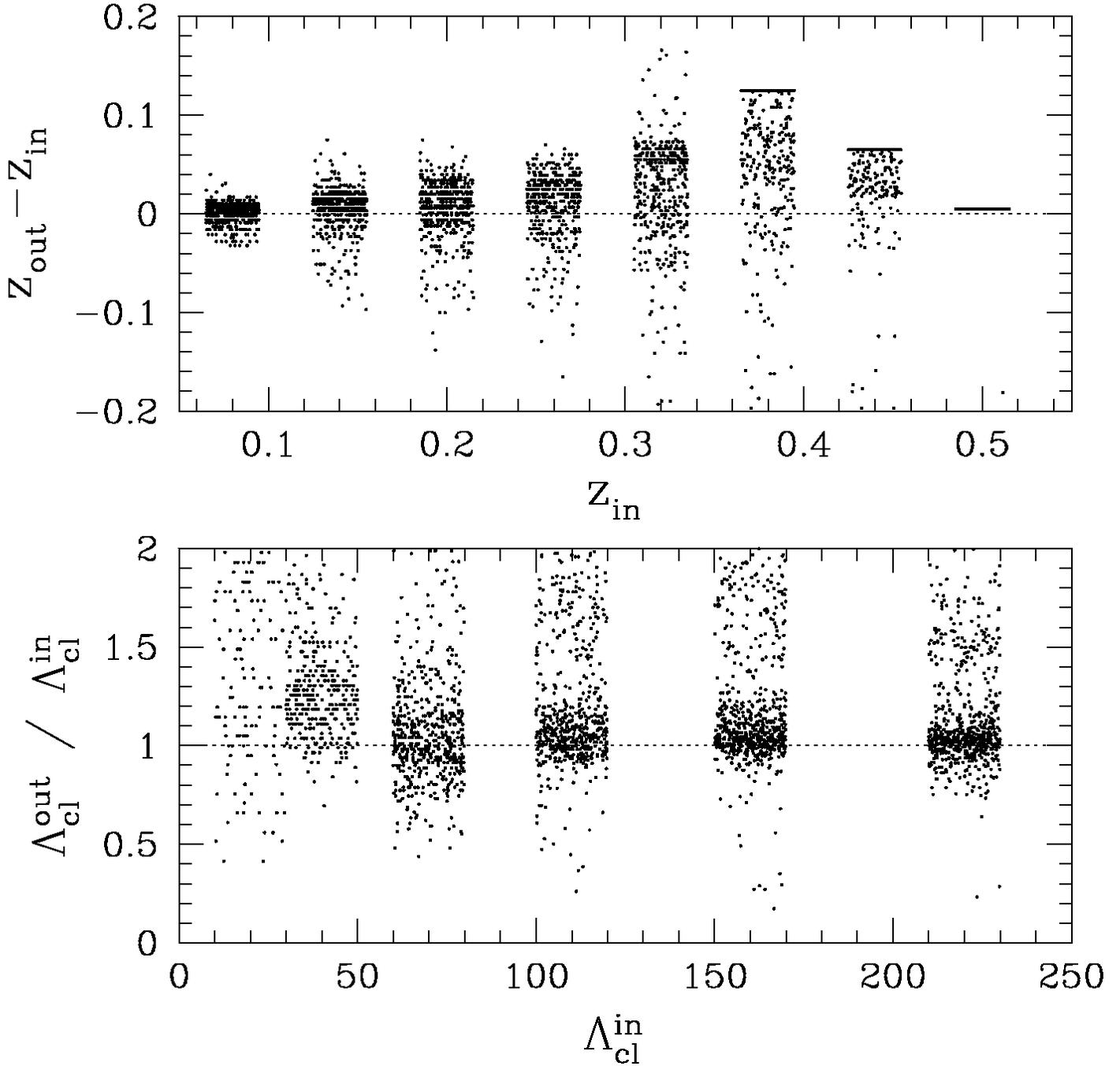}{5.8in}{0}{90}{90}{-290}{-90}
\caption{The input and output parameters evaluated by
the HMF (equivalent to the AMF fine filter). The input values are
discrete, $\Lambda_{cl} = [ 20, 40, 70, 110, 160, 220 ], \, z = [
0.08, 0.14, 0.20, 0.26, 0.32, 0.38, 0.42, 0.5 ]$, and are therefore
shown with a random scatter along the x-axes with a width of $\delta z
= 0.03$ and $\delta\Lambda_{cl} = 20$, to facilitate visual
identification. 
\label{fig:mc_param}}
\end{figure}

We next investigate how HMF parameter evaluations depend upon the
local background density.  First, we show in Figure~\ref{fig:mc_param}
the input and output values of $z$ and $\Lambda_{cl}$ for all
detections in the Monte Carlo simulation.  The distribution of $\Delta
z$ is somewhat positively skewed, implying an overestimation of
redshifts, which is pronounced as we go to higher $z$: the median
values of $\Delta z$ are 0.007, 0.02 and 0.05 for input redshift
ranges $0<z\leq0.2, 0.2<z<0.35$ and $z> 0.35$, respectively.  Also,
half of the clusters with input redshifts $z\gsim0.4$ are recovered
with $z$ estimates at the upper $z$ limit ($z=0.5$).  This trend was
noted in the Matched Filter algorithm by P96, which usually happens
with weak signals (poor and/or high redshift clusters), or in the
field where there are no clusters, where the number-magnitude
distribution of the background is such that the cluster likelihood
calculated yields a monotonically increasing function with redshift.
This could possibly be remedied by tweaking the input luminosity
function to avoid such a trend in the likelihood function (e.g.,
flatter faint end slope), but this could bias parameter estimation in
other ways, and we will not pursue this here.

We also find that the overestimation of redshift is amplified by the
use of a non-uniform background. We have carried out the experiment of
running the HMF on a given set of simulated clusters in both a uniform
and a clustered background.  We find that using the real background
indeed increases the median redshift by about $10-15\%$, but also
creates a long tail of under-estimated redshifts.  We suggest that
this is partly due to real clusters in the data that intercept the
inserted clusters, thus affecting the output redshifts.  The other
concern pertains to the centroid of recovered clusters. While we may
recover unbiased values exactly at the input center, the recovered
clusters are always somewhat ``offset'' from the real center, and this
could lead to some systematic effects. We thus repeated the Monte
Carlo simulation using the true centers and the recovered centers, and found no
sigificant systematic bias in the richness between the two runs.
However, we do find that the number of clusters falsely estimated to
be at $z=0.5$ was reduced by $\sim 15\%$ with the true centers.

The $\Lambda_{cl}$ input and output values are tightly
correlated and do not show much systematic effects, but there is an
extended tail of clusters with overestimated $\Lambda_{cl}$.  These
are the clusters that were falsely assigned to $z=0.5$, which boosts
the estimated $\Lambda_{cl}$ to make up for the loss of galaxies
beyond the magnitude limit.  In what follows, we have left out all
detections with $z_{est}=0.5$ to avoid such complications.

\begin{figure}[p]
\plotfiddle{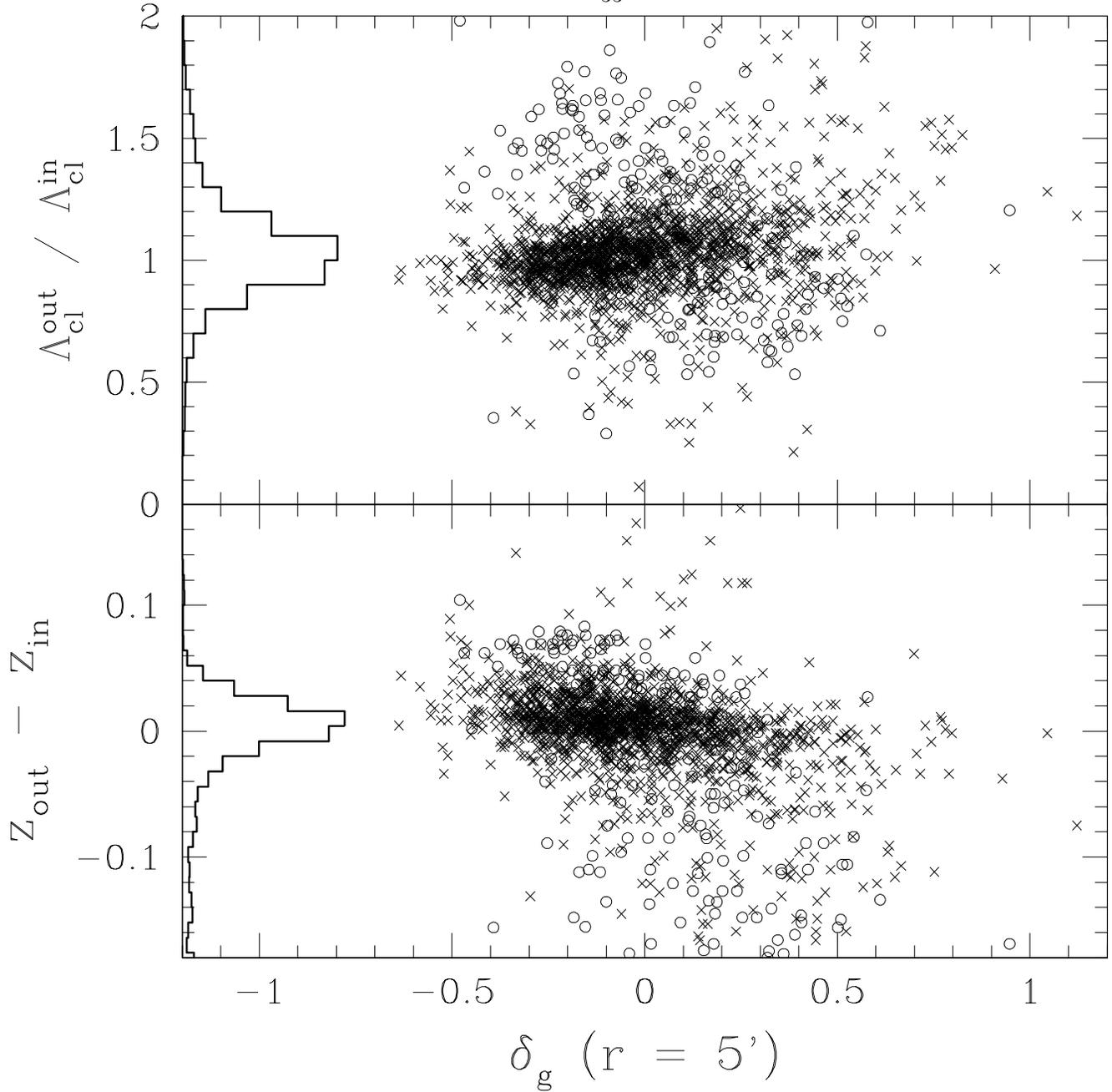}{5.8in}{0}{90}{90}{-290}{-140}
\caption{The dependence of the parameters evaluated by the HMF
(equivalent to the AMF fine filter), on the local background density
contrast. Each point represents one cluster; the difference in the
input and output values of $z$ and the ratio of input and output
$\Lambda_{cl}$ are plotted against $\delta_g$ of the background galaxy
distribution, evaluated within a circle of $r = 5'$ around each
cluster.  The distributions of the points are shown as histograms on
the left. The crosses are clusters which have an input redshift $z
\leq 0.3$, the circles are clusters with input redshift $0.3 < z <
0.5$. The outliers are mostly high redshift clusters whose signals are
weaker.
\label{fig:cc_param}}
\end{figure}

For each cluster recovered by the HMF, Figure~\ref{fig:cc_param} shows
the difference between the input value and recovered value of $z$ and
the ratio of the input and output values of $\Lambda_{cl}$, plotted
against the density contrast of the local background, $\delta_g$.  The
value of $\delta_g$ at each cluster position is evaluated from counts
in cell statistics on the background with a radius of 5 arcmin.  This
radius is chosen to represent the immediate local background of a
cluster ($\sim$ 1 Abell radius at the median redshift of 0.2), much
smaller than what was used for Figure~\ref{fig:cc_det}, $(0.5
\mbox{deg})^2$, where the bin was kept large in order to contain
enough clusters to do statistics.  The figure shows a slight
correlation between the parameters and the local background density
contrast.  
This is to some degree expected, since the model assumes a uniform
background, therefore, a higher background density results in a higher
estimation of richness $\Lambda_{cl}$, and vice versa.  The redshift
dependence shown in the lower panel of Figure~\ref{fig:cc_param} is
due to the way in which the evaluation of the two parameters $z$,
$\Lambda_{cl}$ are correlated.  There are two effects operating.
If the redshift is overestimated, then the angular extent of the
cluster ($r_{\mbox{max}}$), within which $\Lambda_{cl}$ is calculated,
is underestimated, resulting in a underestimated value of
$\Lambda_{cl}$.  However, if the redshift is overestimated, the value
of $\Lambda_{cl}$ is over-compensated for the loss of galaxies fainter
than the survey limit, but this effect is only significant near the
faint limit of the survey (at the high-$z$ limit).  Since the majority
of recovered clusters are those at lower redshifts, the results
exhibit an anti-correlation between the evaluated $z$ and
$\Lambda_{cl}$.  Figure~\ref{fig:cc_param} also demonstrates that most
of the outliers in the parameter evaluations are high redshift
clusters ($z>0.3$), shown as circles in the scatter plot.

\section{Discussion and Summary}
\label{sec:discussion}
We have presented a comparison of three cluster finding
algorithms which are being used to define a cluster catalog from
commissioning data of the Sloan Digital Sky Survey. 
The three algorithms are two Matched
Filter Algorithms (MF and AMF), and the Voronoi Tessellation
Technique (VTT) which is introduced in its
current form in this paper. By applying both Matched Filters on the same galaxy
distribution, we have found that the MF is more efficient in locating
the clusters, whereas the AMF evaluates the cluster parameters
more accurately. This has motivated us to put forward a hybrid  
method (HMF) which uses the MF to select clusters and the AMF to evaluate
the redshifts and the richnesses for those clusters.

The MF is more efficient in selecting clusters than the AMF is because
its thresholding method is redshift dependent.  The AMF locates peaks
in redshift space first, and selects candidates with a signal above a
prescribed threshold regardless of the redshift.  On the other hand,
the MF selects cluster candidates from peaks in the likelihood map of
each assumed redshift; those lying above a threshold that is newly
determined from each map.  As similar clusters at different redshifts
will have very different signals (weakening as one goes to higher
$z$), it is not surprising to find that a redshift-dependent cut
results in a better performance.  Hence in future work, the AMF should
be modified to adopt the peak selection procedure of the MF for better
performance, in order to make use of the further advantages in the AMF
(e.g., using three dimensional information and a parallelization scheme; 
Kepner \& Kim~2000).

In the VTT method we have applied a filter in color-magnitude space to
select galaxies that are most likely members of clusters at a certain
redshift.  This greatly enhances their contrast relative to the
background.  This idea is in principle yet another example of a {\it
matched filter}; this time in color-magnitude space, although it is
not a maximum likelihood method like the MF or the AMF.  There are a
number of existing algorithms that efficiently use this
color-magnitude relation as a filter, but are more restrictive: the
red-sequence method of Yee \& Gladders (1999; Gladders \& Yee 2000)
and the maxBCG technique (also applied to the SDSS) of Annis
\etal~(2001).  While using a restrictive color-magnitude relation
could enhance the efficiency quite a bit, it is also likely to suffer
from selection biases, such as missing clusters with significant blue
populations of galaxies, namely, the Butcher-Oemler (1984) clusters.

Our current C-M filter for the VTT is generous, so that we can focus
on the differences from the Matched Filter algorithms, where a
specific cluster model is used.  These differences will be further
discussed in Paper II with real clusters; in the current paper we
investigate the performance of this new method only with simulated
clusters that exactly follow the spatial and luminosity profile
assumed by the Matched Filter. Despite this advantange for the MF, the
VTT has a similar selection function and better false positive rate
compared to the MF, which points to the power of the technique, and
suggests
that photometric redshifts (which use color information)
will significantly improve the performance of the AMF. 

A Monte Carlo test for the HMF and the VTT was carried out with
simulated clusters inserted in $25 \mbox{deg}^2$ of SDSS background.
We found that the HMF shows a larger drop in detection efficiency in
the presence of a non-uniform background than does the VTT.  This
effect may be due not only to the non-uniformity, but also to
overlapping foreground and background clusters in the data, which can
cause some of the inserted clusters to be overlooked.  This effect is
roughly 15\% for Abell richness class $\geq 0$ clusters in the
redshift range of our interest, if we assume a random distribution of
clusters.  The VTT uses color information (although generous) that
provides stability against such contamination, while the HMF assumes a
uniform background and hence is affected more by this
non-uniformity. Thus using a proper model of the background as a
function of position can further improve the efficiency of the HMF
(Lobo \etal~2000).

We have determined appropriate detection thresholds for the final
cluster catalog that is to be drawn from the SDSS data itself.
These thresholds are: $\sigma_{cut} = 5.2$ and $N_{cut} = 9$ for the HMF and
the VTT respectively.  These values give a reasonable recovery
fraction; only 15\% of clusters detected in the uniform background
case are not recovered with the realistic background, while the lower
limit on false detections, being a small fraction of the expected surface
density of real systems, is acceptable ($1 \mbox{deg}^{-2}$ for the HMF 
and $0.56 \mbox{deg}^{-2}$ for the VTT).

The selection functions for both algorithms were evaluated using these
detection thresholds.  The performance of both methods are very
similar, although the VTT efficiency tends to drop in the intermediate
redshift range compared to the HMF, and is slightly better for lower
redshift.  Both methods are complete for rich clusters ($\Lambda_{cl}
\geq 70$) up to $z \gsim 0.4$.  We compare our selections functions to
those of BNP00, where a similar cluster finding technique was used to
find clusters in shallower data. We find that the performances are
similar for the very low richness clusters ($\Lambda \sim 20$), while
the SDSS outperforms BNP00 by going deeper for the richer clusters
($\Lambda \geq 40$).


Finally, we have shown that the detection efficiencies of both the HMF
and the VTT are nearly independent of the local density of the
background, while the estimated redshift and richness from the HMF are
only slightly biased as a function of the local background density.


In Paper II, we present the cluster catalog compiled from the SDSS
data using these two methods, the HMF and the VTT.  With this we will
be able to test various properties of the algorithms using real
clusters.  Photometric redshifts for the SDSS, soon available, will
significantly improve the AMF in particular.  In future work, it
should be very interesting to compare the above methods with
other existing techniques, e.g., the maxBCG technique (Annis
\etal~2001) that uses far more restrictive color-magnitude information
than the VTT, Cut and Enhance (Goto \etal~2001) that uses the proximity in 
color-space as an enhancement method,  
or the Expectation Maximization algorithm (Nichol
\etal~2000), which does not impose any model constraints at all.
Different algorithms show different results mainly for the fainter
clusters -- poor or distant clusters -- the regime where clusters are
abundant but our understanding is poor.  Their detection will
inevitably depend on different aspects of the method that is used.  If
one were to explore the properties of cluster parameters such as the
luminosity function or the density profile, using the Matched Filter
that constrains these models {\it a priori} will be highly
inappropriate; when exploring the density-morphology relation
of clusters, one should avoid using any color constraints in their
selection.  Therefore, there will be no one technique ideal for all
aspects of cluster science; 
each cluster catalog should be accompanied with a proper
understanding of the nature of the detection method, so that it can be
used for appropriate cluster studies.

\acknowledgements

The Sloan Digital Sky Survey (SDSS) is a joint project of The
University of Chicago, Fermilab, the Institute for Advanced Study, the
Japan Participation Group, The Johns Hopkins University, the
Max-Planck-Institute for Astronomy (MPIA), the Max-Planck-Institute
for Astrophysics (MPA), New Mexico State University, Princeton
University, the United States Naval Observatory, and the University of
Washington. Apache Point Observatory, site of the SDSS telescopes, is
operated by the Astrophysical Research Consortium (ARC).  Funding for
the project has been provided by the Alfred P. Sloan Foundation, the
SDSS member institutions, the National Aeronautics and Space
Administration, the National Science Foundation, the U.S. Department
of Energy, the Japanese Monbukagakusho, and the Max Planck
Society. The SDSS Web site is http://www.sdss.org/.  We thank the
referee, Chris Collins, as well as Donald P. Schneider and Martin
Kerscher for very useful comments on the manuscript.  We also thank
David Spergel for discussion and help on RSJK's thesis, of which this
paper is a part.  RSJK and MAS acknowledge the support of NSF grant
AST96-16901 and AST-0071091 and the Princeton University Research
Board, RSJK acknowledges the support of grants NSF AST98-02980 and 
NASA LTSA NAG5-3503.


\clearpage

\begin{table*}
\caption{Summary of the Procedures for the Two Matched Filter Algorithms}
\begin{small}
\begin{center}
\begin{tabular}{|c|l|l|} \hline \hline
 &   &  \\
Step & Original Matched Filter (MF)&Adaptive Matched Filter (AMF) \\ 
 &   &  \\
\hline \hline
1 & Input galaxy catalog [position, magnitude]  & same \\ 
\hline
2 & Evaluate ${\cal L}_{\mbox{coarse}}$ on a uniform grid for assumed $z_i$ &
Calculate ${\cal L}_{\mbox{coarse}}$ on a $z$-grid for a galaxy position $x_i$ \\
\hline
3 & Repeat step 2 for $i=0,n$ ($n=$ \# of $z$ bins) & 
Repeat step 2 for $i=0,m$ ($m=$ \# of galaxies) \\ 
\hline
4 & Save all likelihood maps for all $z_i$'s  & 
Save $z_{\mbox{coarse}}$ and  ${\cal L}_{\mbox{coarse}}$ where  ${\cal L}_{\mbox{coarse}} (z)$ is \\
 & & maximum, for all $x_i$'s \\ 
\hline
5 & Calculate $\sigma$ for each local maximum within each & \\
  & map, register cluster candidates if $\sigma > \sigma_{\mbox{cut}}$ & \\
\hline
6 & Combine all candidates from all $z_i$ maps, filter &  
Find $x_i$ with highest ${\cal L}_{\mbox{coarse}}$, register as a cluster, \\
  & overlaps and define $z_{\mbox{est}}$ for each cluster &
eliminate nearby $x_i$'s, repeat until ${\cal L}_{\mbox{coarse}} < {\cal L}_{\mbox{cut}}$ \\
\hline 
7 & & Rerun fine filter on cluster positions from step 6, \\
  & & determine $z_{\mbox{fine}}$ and $\Lambda_{\mbox{fine}}$  \\
\hline
  & Final Product : Cluster positions, $z_{\mbox{est}}$, $\Lambda_{cl}$, $\sigma_{\mbox{det}}$ &
Final Product : Cluster positions, $z_{\mbox{fine}}$, $\Lambda_{\mbox{fine}}$, ${\cal L}_{\mbox{fine}}$ \\
  & using $\sigma_{\mbox{det}} > \sigma_{\mbox{cut}}$ & using ${\cal L}_{\mbox{coarse}} > {\cal L}_{\mbox{cut}}$ \\
\hline

\end{tabular}
\label{tab:mf}
\end{center}
\end{small}
\end{table*}

\end{document}